\documentclass[trackchanges]{aastex701}

\usepackage{amsmath}
\usepackage{float}
\usepackage{booktabs}

\received{XXX, 2026}
\revised{XXX, 2026}
\accepted{XXX, 2026}
\begin{document}

\title{Synchrotron Self-Absorption Spectral Modeling Reveals a Magnetically Driven Shock-in-Jet Scenario in Blazar 1156$+$295}

\correspondingauthor{Lang Cui}
\shortauthors{Xu et al.}

\author[orcid=0009-0005-8912-4804,sname='Xu']{Wancheng Xu}
\affiliation{State Key Laboratory of Radio Astronomy and Technology, Xinjiang Astronomical Observatory, Chinese Academy of Sciences, 150 Science 1-Street, Urumqi 830011, China}
\affiliation{School of Astronomy and Space Science, University of Chinese Academy of Sciences, No.1 Yanqihu East Road, Beijing 101408, China}
\affiliation{Konkoly Observatory, HUN-REN Research Centre for Astronomy and Earth Sciences, Konkoly Thege Miklós út 15-17, H-1121 Budapest, Hungary}
\email{xuwancheng@xao.ac.cn}

\author[orcid=0000-0003-0721-5509,sname='Cui']{Lang Cui}
\affiliation{State Key Laboratory of Radio Astronomy and Technology, Xinjiang Astronomical Observatory, Chinese Academy of Sciences, 150 Science 1-Street, Urumqi 830011, China}
\affiliation{School of Astronomy and Space Science, University of Chinese Academy of Sciences, No.1 Yanqihu East Road, Beijing 101408, China}
\email[show]{cuilang@xao.ac.cn}

\author[orcid=0000-0003-4341-0029,sname='An']{Tao An}
\affiliation{State Key Laboratory of Radio Astronomy and Technology, Shanghai Astronomical Observatory, Chinese Academy of Sciences, 80 Nandan Road, Shanghai 200030, China}
\affiliation{School of Astronomy and Space Science, University of Chinese Academy of Sciences, No.1 Yanqihu East Road, Beijing 101408, China}
\email{antao@shao.ac.cn}

\author[orcid=0000-0003-3079-1889,sname='Frey']{S\'andor Frey} 
\affiliation{Konkoly Observatory, HUN-REN Research Centre for Astronomy and Earth Sciences, Konkoly Thege Miklós út 15-17, H-1121 Budapest, Hungary}
\affiliation{CSFK, MTA Centre of Excellence, Konkoly Thege Miklós út 15-17, H-1121 Budapest, Hungary}
\affiliation{Institute of Physics and Astronomy, ELTE Eötvös Loránd University, Pázmány Péter sétány 1/A, H-1117 Budapest, Hungary}
\email{frey.sandor@csfk.org}

\author[orcid=0000-0001-8221-9601,sname='Wang']{Xin Wang}
\affiliation{State Key Laboratory of Radio Astronomy and Technology, Xinjiang Astronomical Observatory, Chinese Academy of Sciences, 150 Science 1-Street, Urumqi 830011, China}
\affiliation{School of Astronomy and Space Science, University of Chinese Academy of Sciences, No.1 Yanqihu East Road, Beijing 101408, China}
\email{wangxin2019@xao.ac.cn}

\author[orcid=0000-0001-9321-6000,sname='Liu']{Yuanqi Liu}
\affiliation{State Key Laboratory of Radio Astronomy and Technology, Shanghai Astronomical Observatory, Chinese Academy of Sciences, 80 Nandan Road, Shanghai 200030, China}
\email{yuanqi@shao.ac.cn}

\author[orcid=0000-0002-8684-7303,sname='Chang']{Ning Chang}
\affiliation{State Key Laboratory of Radio Astronomy and Technology, Xinjiang Astronomical Observatory, Chinese Academy of Sciences, 150 Science 1-Street, Urumqi 830011, China}
\email{changning@xao.ac.cn}

\author[orcid=0000-0002-1908-0536,sname='Chen']{Liang Chen}
\affiliation{State Key Laboratory of Radio Astronomy and Technology, Shanghai Astronomical Observatory, Chinese Academy of Sciences, 80 Nandan Road, Shanghai 200030, China}
\email{chenliang@shao.ac.cn}

\author[orcid=0000-0001-8256-8887,sname='Zhang']{Yingkang Zhang}
\affiliation{State Key Laboratory of Radio Astronomy and Technology, Shanghai Astronomical Observatory, Chinese Academy of Sciences, 80 Nandan Road, Shanghai 200030, China}
\email{ykzhang@shao.ac.cn}

%\collaboration{all}{The Terra Mater collaboration}

%% Use the \collaboration command to identify collaborations. This command
%% takes an optional argument that is either a number or the word "all"
%% which tells the compiler how many of the authors above the command to
%% show. For example "\collaboration[all]{(DELVE Collaboration)}" wil include
%% all the authors above this command.
%%
%% Mark off the abstract in the ``abstract'' environment. 
\begin{abstract}
Unveiling the launching and driving mechanisms of powerful jets in active galactic nuclei (AGNs) is crucial for understanding the co-evolution of supermassive black holes (SMBHs) and their host galaxies. 1156+295 is a blazar at a redshift of $z=0.729$ and exhibits significant variability in long-term radio monitoring. Using multi-frequency Effelsberg single-dish flux density data from 2007 to 2012, we performed synchrotron self-absorption (SSA) spectral modeling and extracted the turnover frequency and turnover flux density. By combining SSA spectral modeling with the core size and brightness temperature from quasi-simultaneous very long baseline interferometry (VLBI) images, we estimated the jet magnetic-field strength and magnetic flux, and investigated their temporal evolution in 1156+295. The evolution of radio flux density, spectral shape, and jet structure is consistent with the shock-in-jet framework. The inferred magnetic flux reaching or exceeding the magnetically arrested disk (MAD) threshold, together with evidence that magnetic energy release precedes the radio flares, supports a magnetically driven jet scenario. Overall, our results place magnetic-field measurements, spectral evolution, and inner-jet structural changes on a common timeline, providing observational constraints on their coupled evolution during flares.

%This study integrates the evolution of radio flux density, spectral properties, jet structure, and magnetic field, offering a qualitative interpretation into the interdependencies of these parameters through the magnetically driven shock-in-jet model.
%Moreover, this work represents a methodological attempt to investigate AGN jet magnetic fields through the synergistic use of single-dish and VLBI data.
\end{abstract}

%% Keywords should appear after the \end{abstract} command. 
%% The AAS Journals now uses Unified Astronomy Thesaurus (UAT) concepts:
%% https://astrothesaurus.org
%% You will be asked to selected these concepts during the submission process
%% but this old "keyword" functionality is maintained in case authors want
%% to include these concepts in their preprints.
%%
%% You can use the \uat command to link your UAT concepts back its source.
\keywords{
\uat{Blazars}{164} ---
\uat{Active galactic nuclei}{16} --- 
\uat{Interstellar synchrotron emission}{856} ---
\uat{Relativistic jets}{1390} --- 
\uat{Magnetic fields}{994}
}

\section{Introduction} \label{section1}
Blazars, a subclass of active galactic nuclei (AGNs), are among the most energetic objects in the Universe. They are characterized by relativistic jets aligned at very small angles to the observer's line of sight \citep{1995PASP..107..803U}. This class consists of flat-spectrum radio quasars (FSRQs) and BL Lacertae objects (BL Lacs), which are distinguished by their optical emission-line features. The central engine is believed to be a supermassive black hole (SMBH), with masses ranging from $10^6$--$10^{10}$$M_{\odot}$, accreting matter and powering highly collimated, relativistic jets \citep{2019ARA&A..57..467B}. These jets emit synchrotron radiation produced by relativistic electrons gyrating in magnetic fields \citep{2012rjag.book.....B}, which play a key role in jet formation, collimation, and acceleration \citep{1977MNRAS.179..433B,1982MNRAS.199..883B}. Blazars exhibit extreme variability across all wavelengths, significant optical and radio polarization changes, and apparent superluminal motion in their jets \citep{2005AJ....130.1418J,2008A&A...485...51H,2008Natur.452..966M}.

Understanding what controls the launching of relativistic jets from accreting black holes is a long-standing issue in relativistic astrophysics. The prevailing theoretical picture invokes magnetic fields to extract rotational energy either from a spinning black hole  \citep{1977MNRAS.179..433B} or the inner accretion disk \citep{1982MNRAS.199..883B} to power the jets. In recent years, general relativistic magnetohydrodynamic (GRMHD) simulations have supported this picture by showing that rapidly rotating black holes can launch stable, high-Lorentz-factor jets via the Blandford--Znajek (BZ) mechanism \citep{2011MNRAS.418L..79T,2020MNRAS.494.3656L}. This requires sufficient magnetic flux to accumulate near the event horizon, either through advection from larger scales or via in situ generation by a dynamo process. Moreover, shocks can compress and partially order magnetic fields in jets, while turbulence may lead to more tangled configurations \citep{2008Natur.452..966M}. Therefore, investigating the magnetic field properties in blazar jets is crucial for constraining the physical mechanisms behind magnetically driven jet launching, collimation, and particle acceleration. 

Previous studies have estimated magnetic field strengths and compared the inferred magnetic flux with the magnetically arrested disk (MAD) threshold, using either statistical blazar samples \citep[e.g.,][]{2014Natur.510..126Z} or multi-epoch analyses of individual sources \citep[e.g.,][]{2024A&A...685L..11G}. However, despite these advances, a time-domain picture that links the evolution of the magnetic field properties, the radio spectral evolution, and the structural changes in the inner jet is still limited. Filling this gap is key to understanding when and how magnetic energy drives particle acceleration and radiation during flares. Very long baseline interferometry (VLBI) can resolve the inner jets of blazars and provide the core size and kinematic constraints needed to link spectral variability to the physical scale of the emitting region \citep[e.g.,][]{2023ApJ...949...20Y}. Meanwhile, dense multi-frequency single-dish monitoring traces the long-term evolution of the flux density and radio spectrum during flares, including the synchrotron self-absorption (SSA) turnover \citep[e.g.,][]{2024A&A...685L..11G}. Together, these observations place magnetic-field measurements, spectral evolution, and inner-jet structural changes on a common timeline. To provide such a time-domain link, we combine multi-frequency total-flux spectral monitoring with quasi-simultaneous multi-frequency VLBI measurements of the inner-jet structure and kinematics, enabling time-resolved estimates of the jet magnetic-field strength (and magnetic flux) and linking these diagnostics to the observed spectral and structural evolution during flares.

Several methods based on radio observations can be used to reveal and investigate the magnetic field properties of blazar jets, including polarimetric VLBI measurements \citep[e.g.,][]{2017A&A...598A..42H,2018MNRAS.473.1850A} and frequency-dependent core shifts \citep[e.g.,][]{1998A&A...330...79L,2021A&A...652A..14C,2023A&A...672A.130C}. In this work, we focus on the SSA turnover method, which constrains the magnetic field strength using the turnover frequency, turnover flux density, and the size of the emitting region at the turnover \citep{1983ApJ...264..296M}. This technique has been widely applied to blazar jets \citep[e.g.,][]{2017ApJ...841..119L,2018ApJ...852...30A,2020ApJ...902..104L,2021A&A...651A..74K,2022MNRAS.510..815K,2024A&A...685L..11G}.

1156+295 (also known as J1159+2914, 4C +29.45, Ton 599) is a blazar with a redshift of $z=0.729$, exhibiting significant variability across the electromagnetic spectrum from radio to $\gamma$-rays \citep{2004A&A...417..887H}. This unique property has led researchers to conduct long-term monitoring across multiple wavelengths and spatial scales. In radio bands, 1156+295 exhibits multi-periodic flux density variability \citep{2014Ap&SS.352..215L,2014MNRAS.443...58W,2026arXiv260305894S}, which has motivated a range of interpretations. Although supermassive black hole binary (SMBHB) scenarios have been proposed as one possible explanation of quasi-periodic variability \citep[e.g.,][]{2004ApJ...615L...5R,2005AIPC..745..487R}, such variability can also arise from jet-intrinsic processes \citep[e.g.,][]{2017A&A...597A..80H}. One possible explanation for the observed quasi-periodic variability is the shock-in-jet scenario \citep{1985ApJ...298..114M}, which suggests that the jets of black holes undergo Kelvin--Helmholtz instability \citep{1987ApJ...318...78H}. In this case, disturbances in the accretion flow are amplified, resulting in spiral jets and multi-periodic radio flux density variability. This framework can explain the observed jet behavior without requiring a binary origin. We therefore focus on the shock-in-jet interpretation and the associated magnetic-field and spectral evolution during flares in 1156+295.

%One explanation for this phenomenon is the shock-in-jet scenario \citep{1985ApJ...298..114M}, which suggests that the jets of black holes undergo Kelvin--Helmholtz instability \citep{1987ApJ...318...78H}, where disturbances in the accretion flow are amplified, leading to spiral jets and multi-periodic radio flux density variability. Another possibility is that the phenomenon is caused by supermassive binary black holes \citep{1980Natur.287..307B}, where the orbital motion of the black holes disturbs the jets, producing similar multi-periodic flux density variations. Although current observations cannot directly confirm the binary black hole hypothesis, studying long-term radio variability and jet kinematics at milliarcsecond (mas) scales can help reveal the magnetic field properties and the driving mechanisms of the jets, allowing us to distinguish between these two scenarios.

In this paper, we analyze the magnetic field properties and evolution in 1156+295 using SSA spectral modeling, based on five years of multi-frequency single-dish and VLBI radio monitoring data from 2007 to 2012. By connecting the evolution of jet structure and kinematics on mas scales with the long-term radio flux density and spectral variability, we aim to reveal the key structural variations in the inner jet responsible for the spectral changes. Together, these multi-epoch measurements of the magnetic field strength, jet power, and radio spectral morphology are expected to provide key insights into particle acceleration and magnetic field evolution near the central engine. This paper is organized as follows. Section~\ref{section2} describes the data reduction and results. Section~\ref{section3} provides an analysis and discussion. Section~\ref{section4} summarizes our conclusions. We assume a standard Lambda Cold Dark Matter ($\Lambda$CDM) cosmological model with $\Omega_{\rm m} = 0.31$, $\Omega_{\rm \Lambda} = 0.69$, and $H_0 = 67.7$~km\,s$^{-1}$\,Mpc$^{-1}$. In this model, the luminosity distance at $z=0.729$ is $D_{\rm L}=4607~\rm Mpc$, 1~mas angular scale corresponds to a projected linear size of 7.471~pc, and 1~mas\,yr$^{-1}$ apparent proper motion corresponds to $24.3\,c$ apparent speed, where $c$ denotes the speed of light. For the definition of spectral index $\alpha$, we follow the convention $S \propto \nu^{+\alpha}$, where $\nu$ is the frequency and $S$ the flux density.

\section{Data Reduction and Results} \label{section2}
\subsection{Radio Variability}\label{section2.1}

In this study, we focus on multi-epoch and multi-frequency VLBI and Effelsberg single-dish observational data between 2007 and 2012 to investigate the radio variability, spectral evolution, jet kinematics, and magnetic field strength of the blazar 1156+295.
The 100-m Effelsberg single-dish radio telescope measurement data are obtained from the \textit{Fermi}-GST AGN Multi-frequency Monitoring Alliance (F-GAMMA) program \citep{2019A&A...626A..60A}, which was dedicated to the monthly flux-density monitoring of selected \textit{Fermi} Gamma-ray Space Telescope (GST) AGNs from January 2007 to January 2015. The frequency coverage of the Effelsberg data includes eight bands: 2.64, 4.85, 8.35, 10.45, 14.6, 23.05, 32, and 43 GHz.

The VLBI imaging data comprise historical observations at $S$, $X$, $Ku$ and $Q$ bands (around 2.3, 8.4, 15.4, and 43~GHz frequencies, respectively). Among them, the $Ku$-band VLBI data were obtained with the U.S. Very Long Baseline Array (VLBA) from the Monitoring Of Jets in Active Galactic Nuclei with VLBA Experiments (MOJAVE) program\footnote[1]{\url{https://www.cv.nrao.edu/MOJAVE/}} \citep{2018ApJS..234...12L}. The $Q$-band VLBI data were obtained from the VLBA Boston University Blazar Monitoring Program (VLBA-BU-BLAZAR)\footnote[2]{\url{https://www.bu.edu/blazars/VLBAproject.html}} \citep{2017ApJ...846...98J}. The VLBI data at the other bands were derived from various astrometric/geodetic snapshot observations archived in the Astrogeo database\footnote[3]{\url{https://astrogeo.org/} maintained by Leonid Petrov} \citep{2025ApJS..276...38P}.

\begin{figure*}[htbp]
\centering
\includegraphics[width=1\textwidth]{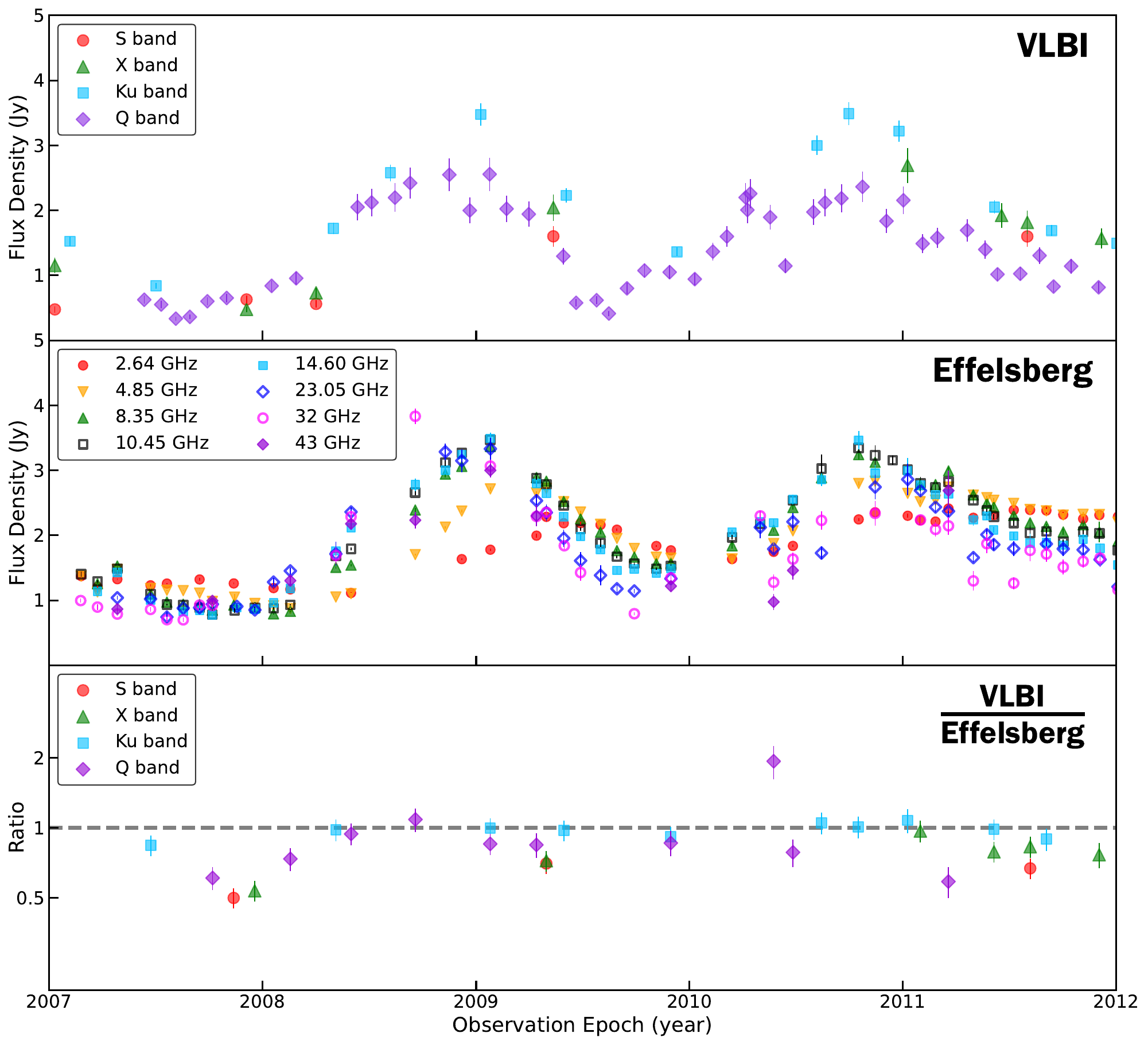}
\caption{Top and middle panels: Radio light curves of 1156+295 constructed using VLBI (top) and Effelsberg single-dish data (middle) during 2007--2012. The same symbols used to denote VLBI bands are applied to the corresponding frequencies on the Effelsberg light curves. Bottom panel: VLBI-to-single-dish flux density ratio at matched epochs.}
\label{figure1}
\end{figure*}

For the calibrated VLBI visibility data, we performed conventional hybrid imaging using iterations of \texttt{CLEAN} deconvolution and self-calibration within the \texttt{Difmap} software package \citep{1997ASPC..125...77S}. The total flux density in each image was then obtained by integrating the flux densities of all {\tt\string CLEAN} components with a signal-to-noise ratio (SNR) exceeding 5. We assume $10\%$ error for the flux densities to account for uncertainties in the VLBI amplitude calibration from antenna system temperatures and gain curves. The VLBI flux densities are listed in Table~\ref{tableA1}. The Effelsberg single-dish flux densities and their uncertainties for 1156+295 were obtained from the publicly available F-GAMMA radio monitoring data published in the Centre de Données astronomiques de Strasbourg (CDS) database \citep{Angelakis2019}. The VLBI and single-dish radio light curves of 1156+295 between 2007--2012 are shown in Figure~\ref{figure1}.

Over the five-year radio monitoring period, 1156+295 exhibits multiple radio flares, as detected in both the VLBI and single-dish observations.
We roughly divide the radio flares during this period into three segments as below:
\begin{itemize}
\item \textbf{2007--2008:} The total flux density exhibits a flat and slightly declining trend, which represents the tail end of a previous radio flare. The examination of earlier radio flux density data indicates that this flare occurred around 2006.
\item \textbf{2008--2010:} A typical radio flare. The flux density reached its peak around 2009 and subsequently declined.
\item \textbf{2010--2012:} The second radio flare. It exhibits a morphology similar to the previous flare between 2008 and 2010, but with a more gradual decay. After the radio flux density decreased to a low level, it remained stable until 2014, although the radio light curves beyond 2012 are not plotted in Figure~\ref{figure1}.
\end{itemize}

During the radio flares, lower-frequency flares consistently lag behind higher-frequency ones, as new jet components originating from the inner regions are first detected at higher frequencies before appearing at lower ones. This behavior in the radio light curves likely reflects the spectral evolution associated with jet launching. 
We matched the flux densities from VLBI and Effelsberg single-dish observations at the same frequency bands within a 30-day window. The calculated ratios locate closely around unity for most epochs (see the bottom panel of Figure \ref{figure1}). Coupled with the similar variability trends exhibited by both datasets, this suggests that the radio emission in blazar 1156+295 is primarily dominated by the compact components imaged with VLBI.
Given the lack of quasi-simultaneous multi-frequency (at least five frequencies) VLBI observations for 1156+295 within a 30-day window, the SSA spectral parameters (turnover frequency $\nu_{\rm m}$ and peak flux density $S_{\rm m}$) derived from single-dish fitting can be used as an approximation for those on VLBI scales.

\subsection{SSA Spectral Fitting}\label{section2.2}

Synchrotron radiation originates from the motion of relativistic electrons in magnetic fields and can be absorbed by electrons in optically thick regions at low frequencies due to energy-level matching. This phenomenon is known as synchrotron self-absorption (SSA). The resulting spectrum peaks where the optical depth is near unity, rising at lower frequencies in the optically thick regime and falling at higher frequencies in the optically thin regime. The shape of SSA spectrum for AGN is defined by the general expression \citep{1999A&A...349...45T}:
\begin{equation}
S_{\nu} = S_\text{m} \left( \frac{\nu}{\nu_\text{m}} \right)^{\alpha_{\text{thick}}} \frac{1-e^{-\tau_\text{m}\left({\nu}/{\nu_\text{m}}\right)^{\alpha_{\text{thin}} - \alpha_{\text{thick}}}}}{1 - e^{-\tau_\text{m}}},
\label{equation1}
\end{equation}
where $S_{\nu}$ represents the flux density at frequency $\nu$, $S_\text{m}$ is the peak flux density at the turnover frequency $\nu_\text{m}$, $\alpha_{\text{thick}}$ and $\alpha_{\text{thin}}$ are the spectral indices in the optically thick and thin parts of the spectrum, located to the left and right of the peak, respectively, and $\tau_\text{m}$ is the optical depth at the turnover frequency $\nu_\text{m}$, satisfying the equation $e^{\tau_\text{m}}-1=(1-\frac{\alpha_\text{thin}}{\alpha_\text{thick}})\tau_\text{m}$, with a third-order approximation given by $\tau_\text{m}\approx \frac{3}{2} \left(\sqrt{1 - \frac{8 \alpha_{\text{thin}}}{3\alpha_{\text{thick}}}}-1 \right)$.

The long-term multi-frequency radio flux density monitoring provides excellent temporal sampling for quantifying spectral evolution and estimating magnetic field strength. 
The observed radio spectra can sometimes be considered as a potential superposition of emission from an underlying quiescent steady-state jet and a perturbed shocked flaring component \citep[e.g.,][]{2011A&A...531A..95F,2022MNRAS.510..815K,2024ApJ...964..176L}. To properly isolate the spectrum of the variable flare component, we employed two distinct methods -- single-epoch and composite-minimum flux density power-law fits -- to extract the quiescent emission spectrum from the historical low-flux-density epochs (late 2007). The results of both methods yield a flat spectral index ($\approx -0.22$), suggesting that the contribution from the quiescent background emission might not be significant. To assess the potential influence of a quiescent background on spectral fitting, we performed a comparative analysis using different modeling approaches, including: (i) the standard single-SSA model; (ii) a residual SSA fit after background subtraction; (iii) a composite fit with a fully fixed background; (iv) a composite fit with a fixed background index but floating normalization; (v) a composite fit with a constrained floating background. For the quiescent background model, we adopted a power-law spectrum $S_{\nu}=1.59\nu^{-0.22}$ derived from a single low-flux-density epoch (2007/07/22). Comparing these fitting results, we found that while explicitly subtracting or fixing the quiescent background (Approaches ii and iii) alters the turnover parameters as expected, allowing the background parameters to float within restricted ranges (Approaches iv and v) generally yields a negligible quiescent contribution, reducing to the single-SSA fit (Approach i). In some cases, the fits even produced inverted spectra for the background, particularly during the rising phase of the flare. These tests suggest that the radio emission of 1156+295 during our observing period is dominated by the variable SSA component. Therefore, to minimize over-parameterization and maintain methodological consistency, we adopt the standard single-SSA model throughout this work, which provides a reliable and physically reasonable description of the data.

Given that the source structure is dominated by mas-scale compact components, and there are no quasi-simultaneous multi-frequency VLBI observations of 1156+295 available, we performed this single-SSA spectral fitting on the total flux densities obtained from the Effelsberg telescope, and used these results instead of model-fitted VLBI flux densities. The SSA model parameters were derived by applying the Bayesian Markov Chain Monte Carlo (MCMC) algorithm with the Python package $\texttt{PyMC}$ \citep{salvatier2016probabilistic}. The input prior distributions for the parameters were defined based on physical constraints, using truncated normal distributions with empirically chosen means and standard deviations. The boundaries of the four parameters in the SSA model were set as follows: $S_\mathrm{m}>0$ Jy, $\nu_\mathrm{m}>0$ GHz, $0<\alpha_\mathrm{thick}<3$, and $-3<\alpha_\mathrm{thin}<0$. The fitting results were evaluated and visualized through posterior predictive checks. Only spectra exhibiting a well-defined peaked profile were selected for subsequent evolution analysis and magnetic field measurements (see Table~\ref{table1} and Figure~\ref{figureD1}).

\subsection{Jet Kinematics, Core Size, and Brightness Temperature}\label{section2.3}

Long-term VLBI observations allow us to study the evolution of AGN core--jet structures on mas scales. To investigate jet structure and kinematics in the blazar 1156+295 between 2007 and 2012, we performed multiple-component elliptical Gaussian model fitting to the VLBI visibility data using the \texttt{MODELFIT} task in \texttt{Difmap} \citep{1997ASPC..125...77S}. The brightest component was considered as the core and other components with SNR exceeding 6 were considered as associated with the jet. In cases where the elliptical Gaussian model fitted to the core exhibited an abnormally small axial ratio -- defined as the ratio of the minor to major axis full width at half-maximum (FWHM) --, we constrained the core component to a circular Gaussian profile instead of an elliptical. In cases where the core component was a point source and thus unresolved (e.g., modeled core major axis was smaller than $1/5$ of the restoring beam minor axis), upper limits equal to one-fifth of the synthesized beam major and minor axes were adopted for the core major and minor axis sizes, respectively \citep[e.g.,][]{2017ApJ...846...98J}.

\begin{figure*}[ht!]
\centering
\includegraphics[width=1\textwidth]{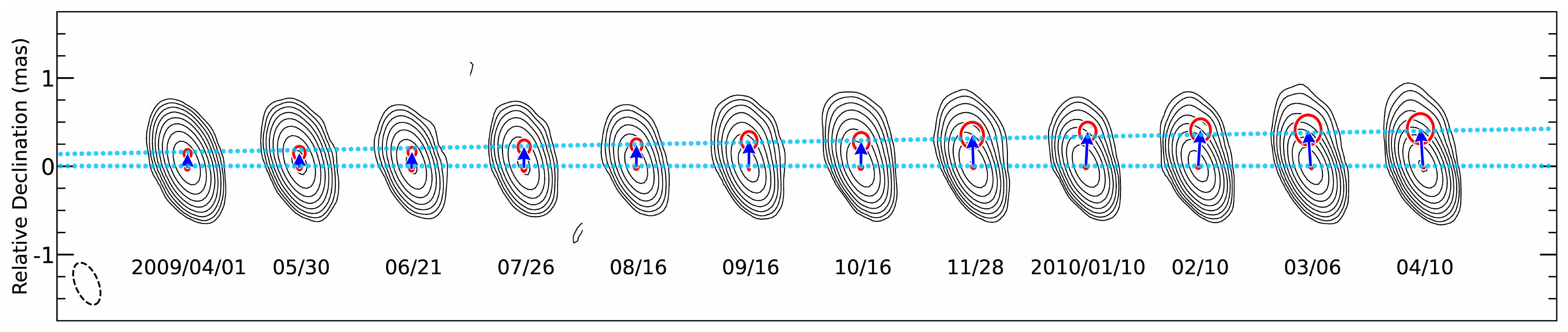}
\caption{The VLBI images at 43 GHz of the blazar 1156+295 convolved with the same beam of $0.52\,\mathrm{mas} \times 0.29\,\mathrm{mas}$ (FWHM) at major axis position angle $29.0^{\circ}$ (dashed ellipse at the lower left) between 2009 April 1 and 2010 April 10. The residual RMS noise is $\sigma=2.34$~mJy\,beam$^{-1}$, and the contours are drawn at $[-1, 1, 2, 4, 8, 16, 32, 64, 128, 256, 512] \times 3\sigma$. The red circles/ellipses represent the positions and sizes of the fitted Gaussian models for the core (central) and jet (upper) components. The horizontal and inclined light blue dotted lines represent the reference position of the core and the fitted trajectory of the moving jet component, respectively. Linear regression provides an apparent jet proper motion of $(0.287 \pm 0.015)$\,mas\,year$^{-1}$ with a coefficient of determination $R^2=0.97$.}
\label{figure2}
\end{figure*}

VLBI imaging reveals a north-directed jet in 1156+295 \citep{2017ApJ...846...98J,2018ApJS..234...12L}. Long-term VLBI monitoring shows significant jet apparent proper motion and the emergence of new inner jet components. The $Q$-band VLBI images were selected for studying jet apparent proper motion and extracting core sizes due to their better time coverage and closer alignment with the epochs of Effelsberg single-dish SSA spectral fitting. The images from 2009 April 1 to 2010 April 10, convolved with the same restoring beam of $0.52\,\mathrm{mas} \times 0.29\,\mathrm{mas}$ (FWHM) at major axis position angle $29.0^{\circ}$, clearly exhibit continuous outward jet component motion (see Figure~\ref{figure2}). It thus allows us to determine the apparent jet speed. Using the measured core--jet separations listed in Table~\ref{tableB2}, the apparent speed of the jet component was estimated via linear fitting to be $\beta_{\text{app}}=(0.287\pm0.015)$~mas\,year$^{-1}$, corresponding to $(7.0\pm0.4)\,c$. 

High-resolution VLBI imaging reveals the size of the compact emission core. To estimate the SSA magnetic field strength at the emission region where the optical depth is unity for turnover frequency $\nu_{\rm m}$, it is necessary to extrapolate the observed core size to $\nu_{\rm m}$. In a conical jet dominated by the synchrotron self-absorbed core, the observed core size is expected to vary as a function of frequency following the relationship \citep{1979ApJ...232...34B,1981ApJ...243..700K,1998A&A...330...79L}:
\begin{equation}
\theta_{\nu} \propto \nu^{-1}.
\label{equation2}
\end{equation}
Based on this relationship and the VLBI-measured $Q$-band core size with an assumed 20\% systematic uncertainty, the core size $\theta_{\rm m}$ at the spectral turnover frequency and its corresponding uncertainty can be estimated.
%where $\epsilon=\frac{1}{1+\alpha_{\rm thick}}$ is the jet geometry parameter.
%and was found to be 0.44 $\pm$ 0.08 \citep{2017ApJ...834...65A,2021A&A...651A..74K}.
%The observed core size $\theta_{\nu}$ at the observational frequency is extracted from the VLBI image and with a uncertainty of:
%\begin{equation}
%\sigma_{\theta_\nu} = \frac{\sqrt{B_{\rm maj} \times B_{\rm min}}}{2 ~ \rm SNR}
%\end{equation}
%where $B_{\rm maj}$ and $B_{\rm min}$ are the major axis and minor axis of the VLBI synthesized beam. SNR is the signal-to-noise ratio of the core component. 
%The uncertainty of the core size at the turnover frequency can be made through error propagation:
%\begin{equation}
%\sigma_{\theta_{\rm m}} = \theta_{\rm m}\sqrt{\left(\frac{\sigma_{\theta_\nu}}{\theta_\nu}\right)^2+ \left(\epsilon \frac{\sigma_{\nu_{\rm m}}}{\nu_{\rm m}}\right)^2+ \left( \ln \left(\frac{\nu_{\rm m}}{\nu} \right) \sigma_{\epsilon} \right)^2}
%\label{formula 9}
%\end{equation}

The Doppler-boosting factor $\delta$ is an important parameter for calculating the SSA magnetic field strength and can be estimated using the ratio between the measured brightness temperature and the intrinsic brightness temperature with the assumption of energy equipartition \citep{1994ApJ...426...51R}:
\begin{equation}
\delta=\frac{T_{\rm b,obs}}{T_{\rm b,int}},
\label{equation3}
\end{equation}
where $T_{\rm b,int}=5\times10^{10} \rm ~K$ represents the intrinsic brightness temperature with the assumption of energy equipartition between radiating particles and the magnetic field in the jet \citep{1994ApJ...426...51R}. $T_{\rm b,obs}$ is the observed brightness temperature which is calculated using \citep{Condon1982}:
\begin{equation}
T_{\rm b,obs} = 1.22 \times 10^{12} (1+z) \frac{S_{\rm \nu}}{\nu^2 \theta_\nu^2} \rm ~[K],
\label{equation4}
\end{equation}
where $\nu$ is the observing frequency in units of GHz, $\theta_{\nu}$ the geometric mean of fitted major and minor axis sizes in mas, and $S_{\nu}$ the core flux density in Jy. In this study, we used the parameters obtained from $Q$-band VLBI imaging data to estimate the observed brightness temperatures and the Doppler factors of 1156+295 core. Key parameters characterizing the jet, the bulk Lorentz factor ($\Gamma$) and the viewing angle with respect to the line of sight ($\theta_\text{view}$), can be calculated from the following formulae, using the apparent jet speed expressed in the units of $c$ ($\beta_{\text{app}}$) and the Doppler factor $\delta$ \citep{1995PASP..107..803U}:
\begin{equation}
\Gamma = \frac{\beta_\text{app}^2+\delta^2+1}{2\delta},
\label{equation5}
\end{equation}
\begin{equation}
{\theta_\text{view}} = \arctan \left(\frac{2\beta_\text{app}}{\beta_\text{app}^2+\delta^2-1}\right),
\label{equation6}
\end{equation}

Through the above series of calculations, we find that for the jet of 1156+295, the Doppler factors, the bulk Lorentz factors, and the jet viewing angles span the ranges of $2.4 \lesssim \delta \lesssim 16.9$, $7.1 \lesssim \Gamma \lesssim 11.6$, and  $2.4^\circ \lesssim \theta_{\mathrm{view}} \lesssim 14.7^\circ$, respectively (see Table~\ref{tableB1}). These parameters are generally consistent with the characteristic features of typical FSRQs, which aligns with the classification of 1156+295 \citep{2009A&A...494..527H}. It is noted that during the epoch around the middle of 2010, when the Doppler factor was particularly high ($\delta>15$) and the viewing angle was small ($\theta_{\rm view}<2.5^{\circ}$), a strong $\gamma$-ray flare was detected \citep{2014MNRAS.445.1636R}, which aligns with the expectation that such a configuration enhances high-energy radiation. It is also possible that during the outbursts, the intrinsic brightness temperature exceeds the equipartition value \citep{2006ApJ...642L.115H,2021ApJ...923...67H}. In such cases, the Doppler factors based on Equation~(\ref{equation3}) are overestimated, and, consequently, the $\theta_\mathrm{view}$ values are underestimated (Equation~\ref{equation6}). 
Previous estimates of the variability Doppler factor for 1156+295 yielded $\delta_{\rm var}\approx 10.85$ around January 2008, and $\delta_{\rm var}\approx 18.54$ around December 2009 \citep{2014MNRAS.445.1636R}. Our values are broadly consistent with these estimates at comparable epochs (see Table~\ref{tableB1}). Moreover, in the SSA magnetic-field expression (Equation~(\ref{equation7}), given below), the Doppler factor enters only to the first power, so its uncertainty is a minor contributor to the overall uncertainty in the inferred magnetic-field strength. The derived core sizes $\theta_{\rm m}$ at the turnover frequency and the observed brightness temperatures $T_{\rm b,obs}$ are summarized in Table~\ref{tableB1}.

\subsection{SSA Magnetic Field Strength}\label{section2.4}

The magnetic field strength can be estimated using the formula derived from the theory of synchrotron emission, which incorporates the turnover frequency ($\nu_{\rm m}$), the peak flux density ($S_{\rm m}$) and the source size ($\theta_{\rm m}$) \citep{1983ApJ...264..296M}:
\begin{equation}
B_{\rm SSA} \approx 10^{-5} b(\alpha) \nu_{\rm m}^{5} {S^{-2}_{\rm m}} \theta_{\rm m}^4 \left(\frac{\delta}{1+z}\right),
\label{equation7}
\end{equation}
where $b(\alpha)$ is a dimensionless factor that depends on the spectral index in the optically thin region and ranges from 1.8 to 3.8 based on the table given in \cite{1983ApJ...264..296M}. In order to estimate $b(\alpha)$, we linearly interpolated the values and fitted for the optically thin spectral indices in SSA spectra derived in Section~\ref{section2.2}. The other parameters in Equation (\ref{equation7}), $\nu_{\rm m}$, $S_{\rm m}$, $\theta_{\rm m}$, and $\delta$, are extracted from SSA spectral fitting and VLBI imaging results. The parameters and results of the SSA magnetic field %feature 
measurements are listed in Table~\ref{table1}. The uncertainties were derived from the model or estimated through error propagation.

\begin{deluxetable}{cccccccccc}[h!]
\tablenum{1}
\tablecaption{Parameters and results of the SSA magnetic field feature measurements in 1156+295.\label{table1}}
\setlength{\tabcolsep}{3.0pt}
\tabletypesize{\footnotesize}
\tablehead{
\colhead{SSA Epoch} & \colhead{\raisebox{-1.5ex}{\shortstack{Matched\\VLBI Epoch}}}
 & \colhead{$\nu_{\rm m}$} & \colhead{$S_{\rm m}$} & \colhead{$\alpha_{ \rm thick}$} & \colhead{$\alpha_{\rm thin}$} & \colhead{$\theta_{\rm m}$} & \colhead{$B_{\rm SSA}$} & \colhead{$B_{\rm 1pc}$} & \colhead{$\Phi_{\rm jet}$}\\ 
\colhead{(yyyy/mm/dd)} & \colhead{(yyyy/mm/dd)} & \colhead{(GHz)} & \colhead{(Jy)} & \colhead{} & \colhead{} & \colhead{(mas$\times$mas)} &  \colhead{(mG)} & \colhead{(G)} & \colhead{(10$^{33}$G cm$^2$)}
} 
\startdata
2007/03/25&2007/06/13&9.72$\pm$0.85&1.22$\pm$0.01&0.04$\pm$0.01&$-$1.20$\pm$0.15&0.38$\times$0.14&18.32$\pm$14.74&1.75$\pm$1.41&4.89$\pm$3.93\\
2007/04/28&2007/06/13&7.66$\pm$0.42&1.54$\pm$0.01&0.19$\pm$0.03&$-$1.10$\pm$0.08&0.48$\times$0.17&9.06$\pm$7.26&1.10$\pm$0.88&3.07$\pm$2.46\\
2008/05/06&2008/06/12&11.81$\pm$0.71&1.66$\pm$0.04&1.29$\pm$0.55&$-$0.66$\pm$0.26&0.28$\times$0.13&$<$37.24&$<$4.08&$<$11.42\\
2008/06/01&2008/06/12&21.10$\pm$0.53&2.11$\pm$0.04&0.45$\pm$0.01&$-$1.19$\pm$0.08&0.15$\times$0.07&$<$46.36&$<$2.84&$<$7.96\\
2008/12/07&2008/12/21&11.47$\pm$0.34&3.26$\pm$0.04&0.62$\pm$0.02&$-$0.95$\pm$0.09&0.50$\times$0.22&29.15$\pm$23.35&4.07$\pm$3.26&11.40$\pm$9.13\\
2009/01/25&2009/01/24&12.10$\pm$0.34&3.49$\pm$0.04&0.87$\pm$0.07&$-$0.42$\pm$0.05&0.44$\times$0.16&14.80$\pm$11.85&1.63$\pm$1.31&4.57$\pm$3.66\\
2009/04/14&2009/04/01&8.62$\pm$0.35&2.89$\pm$0.02&0.89$\pm$0.19&$-$0.37$\pm$0.07&0.40$\times$0.23&$<$4.23&$<$0.52&$<$1.45\\
2009/05/01&2009/04/01&7.26$\pm$0.46&2.81$\pm$0.02&0.95$\pm$0.37&$-$0.30$\pm$0.10&0.47$\times$0.27&$<$3.25&$<$0.47&$<$1.32\\
2009/05/31&2009/05/30&6.25$\pm$0.40&2.55$\pm$0.02&0.66$\pm$0.27&$-$0.44$\pm$0.10&0.35$\times$0.35&2.22$\pm$1.79&0.32$\pm$0.25&0.88$\pm$0.71\\
2009/06/29&2009/06/21&4.81$\pm$0.24&2.36$\pm$0.02&0.53$\pm$0.29&$-$0.53$\pm$0.13&0.74$\times$0.39&$<$1.75&$<$0.60&$<$1.67\\
2009/08/02&2009/07/26&3.77$\pm$0.33&2.19$\pm$0.02&0.45$\pm$0.38&$-$0.50$\pm$0.18&0.94$\times$0.59&2.19$\pm$1.76&1.10$\pm$0.89&3.08$\pm$2.48\\
2010/05/25&2010/05/19&11.01$\pm$0.62&2.08$\pm$0.03&0.15$\pm$0.01&$-$1.25$\pm$0.13&0.32$\times$0.17&$<$29.00&$<$3.11&$<$8.70\\
2010/06/27&2010/06/14&11.78$\pm$0.47&2.50$\pm$0.03&0.25$\pm$0.01&$-$1.19$\pm$0.10&0.29$\times$0.14&$<$11.55&$<$0.97&$<$2.72\\
2010/11/15&2010/12/04&8.92$\pm$0.54&3.10$\pm$0.05&0.37$\pm$0.08&$-$0.72$\pm$0.12&0.38$\times$0.21&$<$6.08&$<$0.70&$<$1.96\\
2011/01/09&2011/01/02&9.21$\pm$0.78&2.86$\pm$0.07&0.26$\pm$0.09&$-$0.79$\pm$0.18&0.37$\times$0.16&8.72$\pm$7.03&1.10$\pm$0.89&3.08$\pm$2.49\\
2011/01/31&2011/02/04&11.01$\pm$0.83&2.71$\pm$0.05&0.14$\pm$0.04&$-$0.73$\pm$0.12&0.43$\times$0.14&12.60$\pm$10.14&1.25$\pm$1.01&3.51$\pm$2.82\\
2011/02/26&2011/03/01&9.23$\pm$0.48&2.75$\pm$0.02&0.27$\pm$0.05&$-$0.71$\pm$0.10&0.56$\times$0.26&18.92$\pm$15.17&3.09$\pm$2.48&8.66$\pm$6.95\\
2011/03/20&2011/03/01&7.25$\pm$0.76&2.88$\pm$0.04&0.54$\pm$0.36&$-$0.50$\pm$0.16&0.72$\times$0.34&11.46$\pm$9.25&2.39$\pm$1.93&6.68$\pm$5.39\\
2011/05/02&2011/04/21&6.59$\pm$0.30&2.69$\pm$0.03&0.28$\pm$0.05&$-$0.98$\pm$0.10&0.55$\times$0.55&12.48$\pm$10.00&3.41$\pm$2.73&9.53$\pm$7.64\\
2011/06/06&2011/06/12&5.03$\pm$0.29&2.54$\pm$0.03&0.65$\pm$0.38&$-$0.52$\pm$0.14&1.84$\times$0.35&9.85$\pm$7.91&4.74$\pm$3.81&13.28$\pm$10.65\\
\enddata
\end{deluxetable}

%The uncertainty of $B_{\rm SSA}$ can be made through error propagation: 
%\begin{equation}
%\sigma_{B_{\rm SSA}} =  B_{\rm SSA}\sqrt{\left(5 \cdot \frac{\sigma_{\nu_{\rm m}}}{\nu_{\rm m}}\right)^2 + \left(4 \cdot \frac{\sigma_{\theta_{\rm m}}}{\theta_{\rm m}}\right)^2 + \left(2 \cdot \frac{\sigma_{S_{\rm m}}}{S_{\rm m}}\right)^2}
%\end{equation}
%Here, the uncertainties of turnover frequency $\sigma_{\nu_{\rm m}}$ and of turnover flux density $\sigma_{S_{\rm m}}$ are extracted from SSA spectral fitting, and it is assumed that the relative error of turnover flux has a similar value at the VLBI imaging scale and the single-dish imaging scale. The uncertainties of core size at the turnover frequency $\sigma_{\theta_{\rm m}}$ are calculated by formula (\ref{formula 9}).

The magnetic field strength $B_{\rm SSA}$ is derived at the location of the opaque core corresponding to the turnover frequency $\nu_{\rm m}$. To provide some practical values at a typical distance, we need to convert the magnetic field strength measured at the location of the opaque core to the value at a standardized distance of 1 pc from the black hole ($B_{\rm 1pc}$). Under the assumption of a narrow, conical jet in equipartition with a constant half-opening angle and Lorentz factor, the magnetic field follows a power-law decay with distance from the central engine \citep{1979ApL....20...15B,1998A&A...330...79L,2005ApJ...619...73H}:
\begin{equation}
B(r) = B_1 \left( \frac{r}{r_1} \right)^{-1},
\label{equation8}
\end{equation}
where $B(r)$ is the magnetic field strength at distance $r$ from the central black hole, $B_1$ is the magnetic field strength at a reference distance $r_1$, and the power-law index of $-1$ is a standard assumption \citep[e.g.,][]{2014Natur.510..126Z,2015MNRAS.447.2726N,2021A&A...652A..14C,2024A&A...685L..11G}.
In this study, we set the reference distance to $r_1=1\,{\rm pc}$, thus $B_{1\rm pc}=B_{\rm SSA}\times r/{\rm 1\,pc}$, where $r$ is the distance in units of pc from the central black hole to the opaque core at $\nu_{\rm m}$. Under the assumption of a conical jet with a small opening angle $\phi_{\rm j}$ and a small viewing angle $\theta_{\rm view}$, the distance $r$ can be geometrically approximated as $r\approx\theta_{\rm m}/\phi_{\rm j}$. Here, $\theta_{\rm m}$ is the core angular size in units of pc at the turnover frequency, and $\phi_{\rm j}$ is the opening angle of the jet. For blazar, the jet opening angle $\phi_{\rm j}$ and the bulk Lorentz factor $\Gamma$ approximately satisfy the relationship $\Gamma\phi_{\rm j} \sim 0.13$ \citep{2021A&A...652A..14C}.

%Therefore, the magnetic field strength at the distance of 1 parsec from the central black hole can be estimated with:
%\begin{equation}
%B_\text{1pc} = B_\text{SSA} \frac{r_\text{c}}{1\text{pc}}
%\end{equation}
%and has uncertainty of $\sigma_{B_{\rm 1pc}}=\sigma_{B_{\rm SSA}} \times r_c$, through the error propagation.
%To convert $B_{\rm SSA}$ to $B_{\rm 1pc}$, we would need to know the distance of the core from the central black hole in parsecs ($r_c$). Assuming the jet has a conical shape with a small half-opening angle and a small viewing angle ($\theta_{\rm view} \lesssim 5.7^\circ$).
%According to the geometric relationship (see Figure~\ref{figure 5}), $r_c$ can be approximately expressed as:
%\begin{equation}
%r_{\rm c} \approx \frac{\theta_{\rm m} \cdot \cos{\theta_{\rm view}}}{2~\tan{\left(\frac{\theta_{\rm j}}{2}\right)}} \approx \frac{\theta_{\rm m}}{\theta_{\rm j}}
%\end{equation}

%The magnetic flux $\Phi_{\rm jet}$ is in units of $\rm [G~cm^2]$, and has uncertainty of:
%\begin{equation}
%\sigma_{\Phi_{\rm jet}} = 1.2 \times 10^{25} f(a_{*}) \Gamma \theta_{\rm j} \left(\frac{M}{M_{\odot}}\right) \sigma_{B_{\rm 1pc}}
%\end{equation}

\section{Discussion}\label{section3}
\subsection{Radio Variability and Spectral Evolution}\label{section3.1}

\begin{figure*}[h!]
\centering
\includegraphics[width=0.32\textwidth]{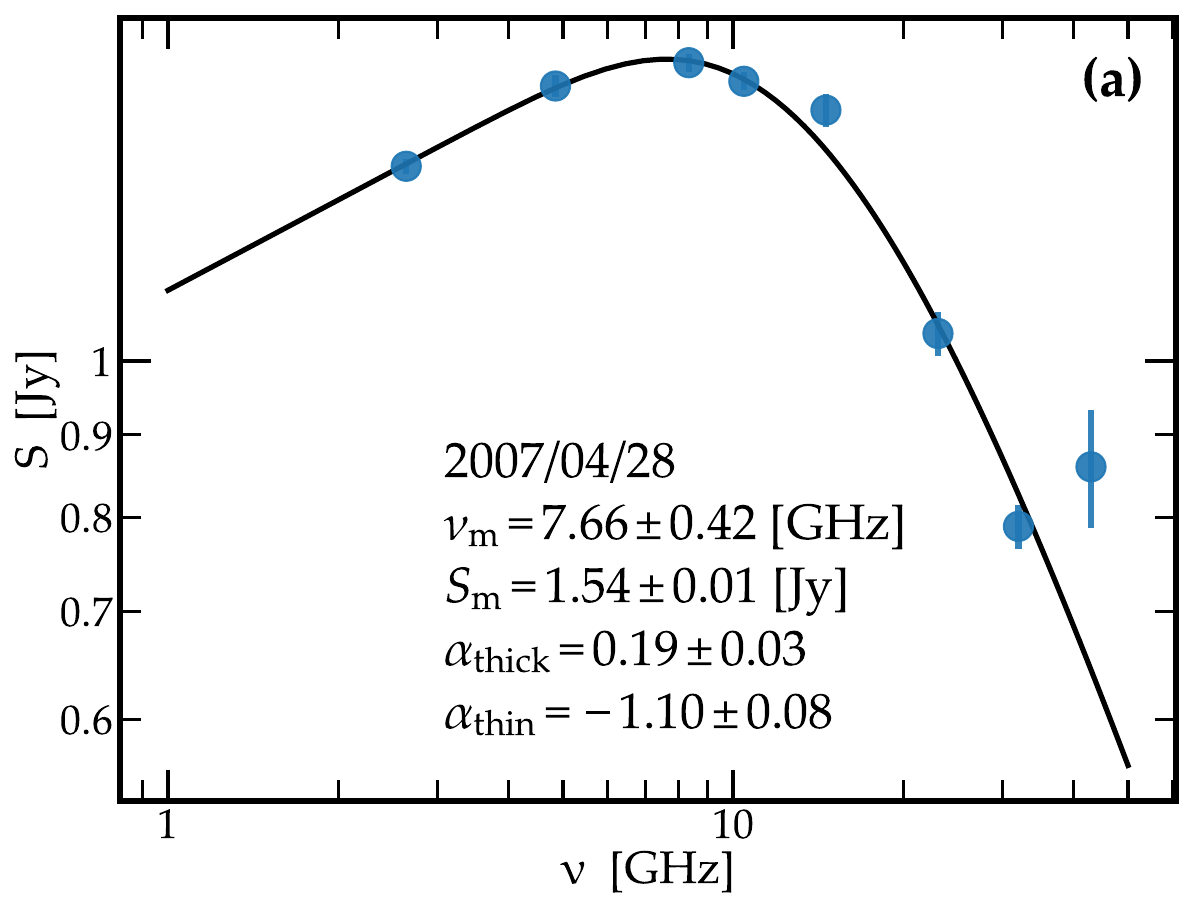}
\includegraphics[width=0.32\textwidth]{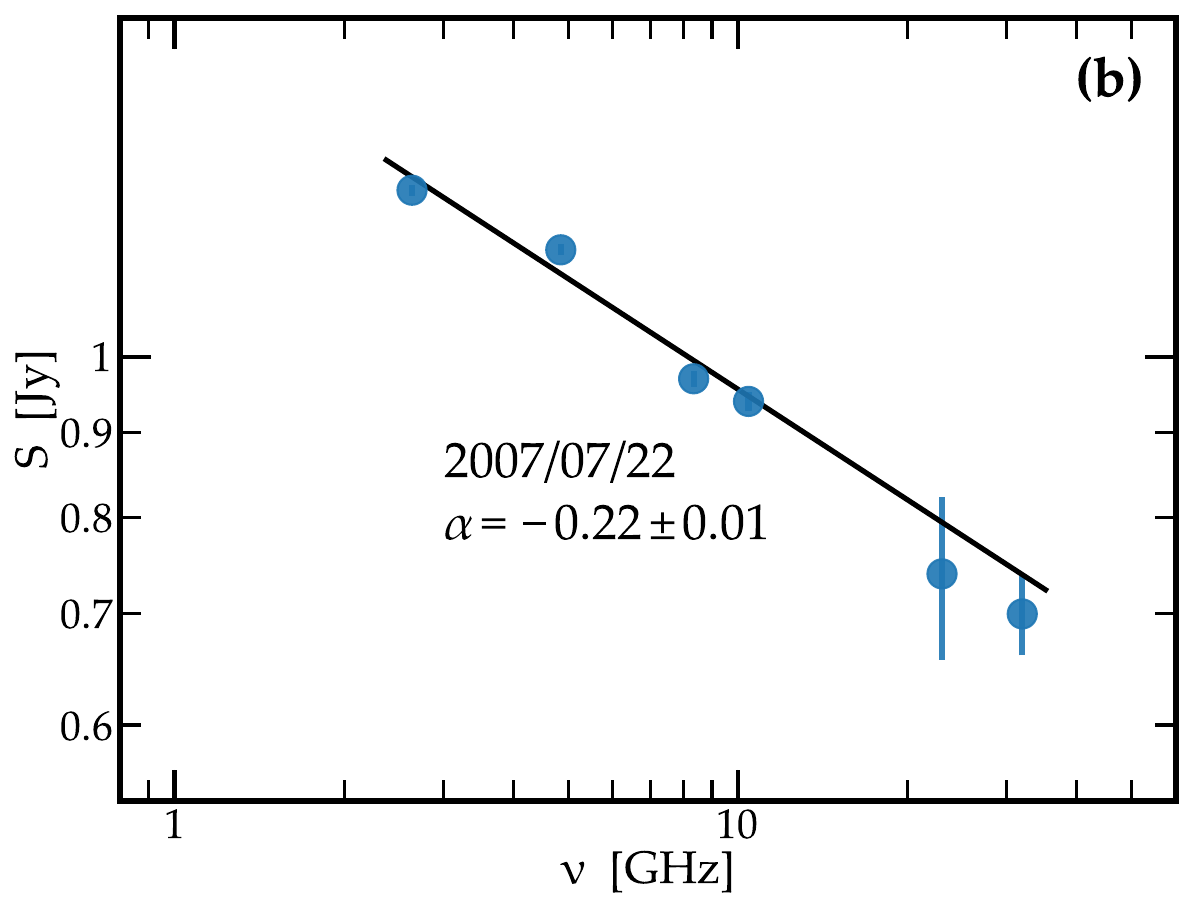}
\includegraphics[width=0.32\textwidth]{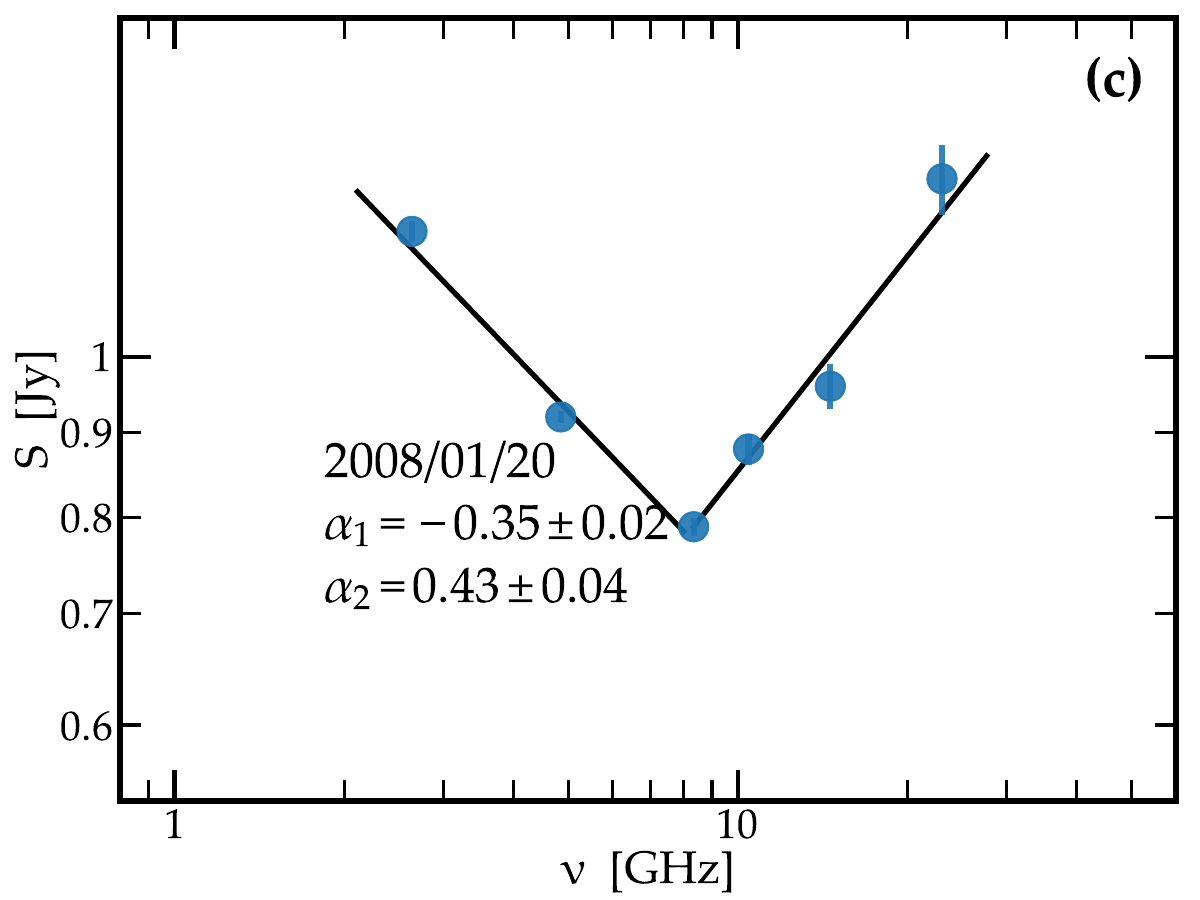}
\includegraphics[width=0.32\textwidth]{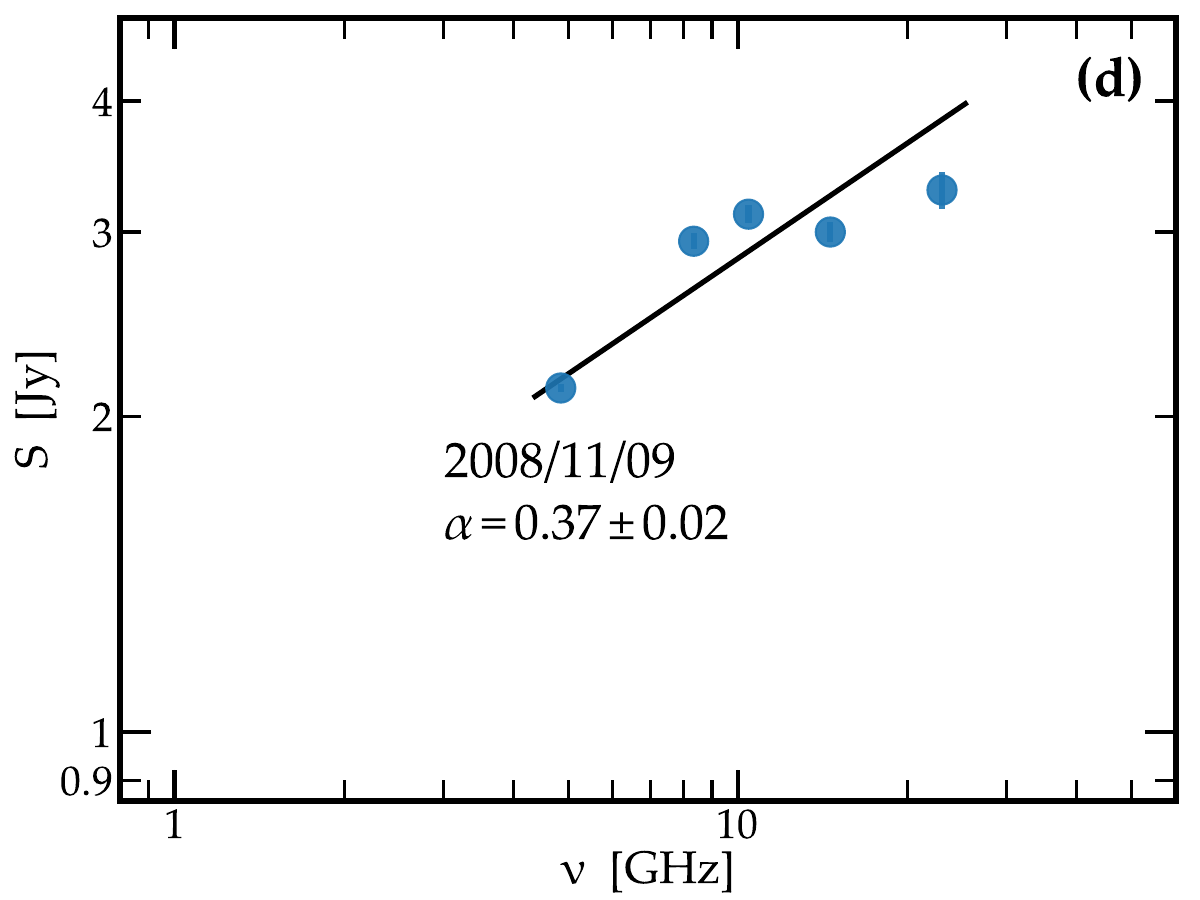}
\includegraphics[width=0.32\textwidth]{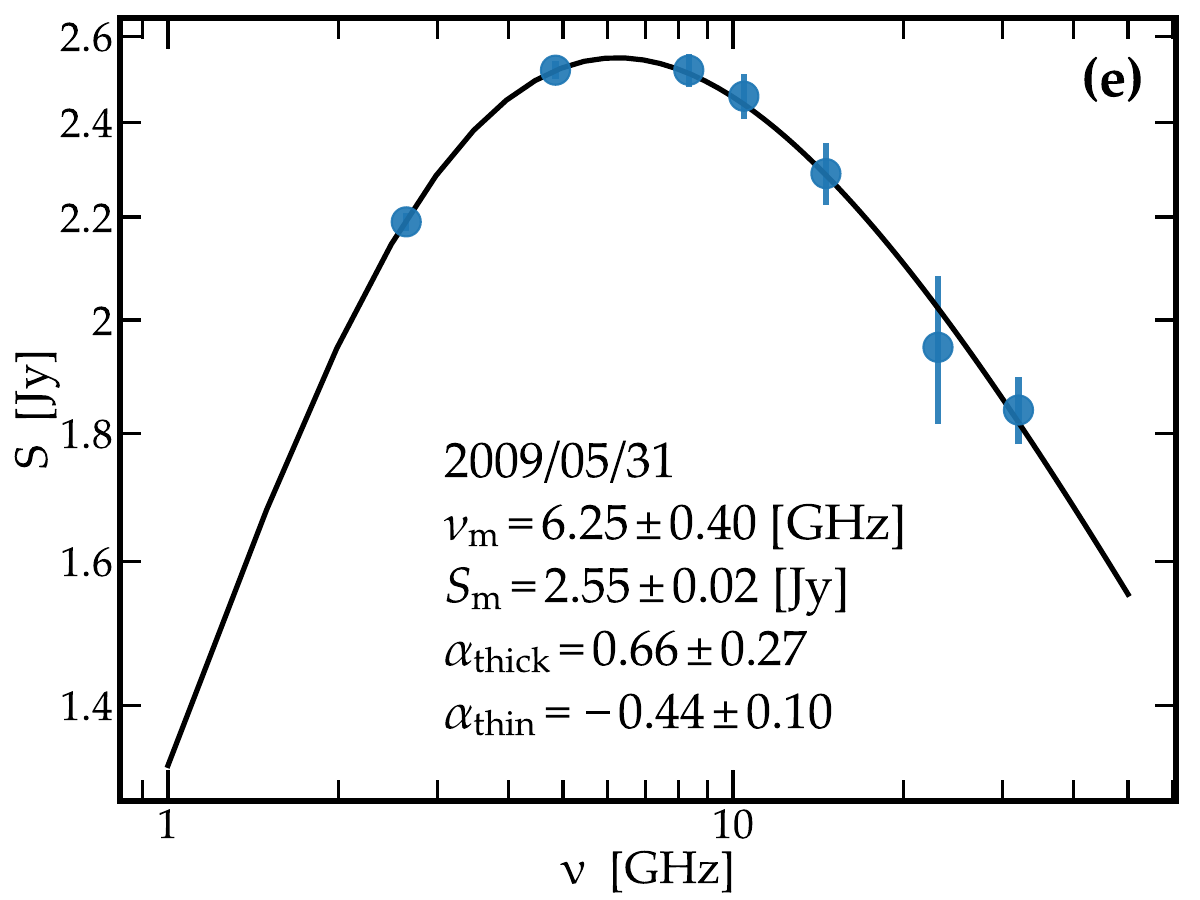}
\includegraphics[width=0.32\textwidth]{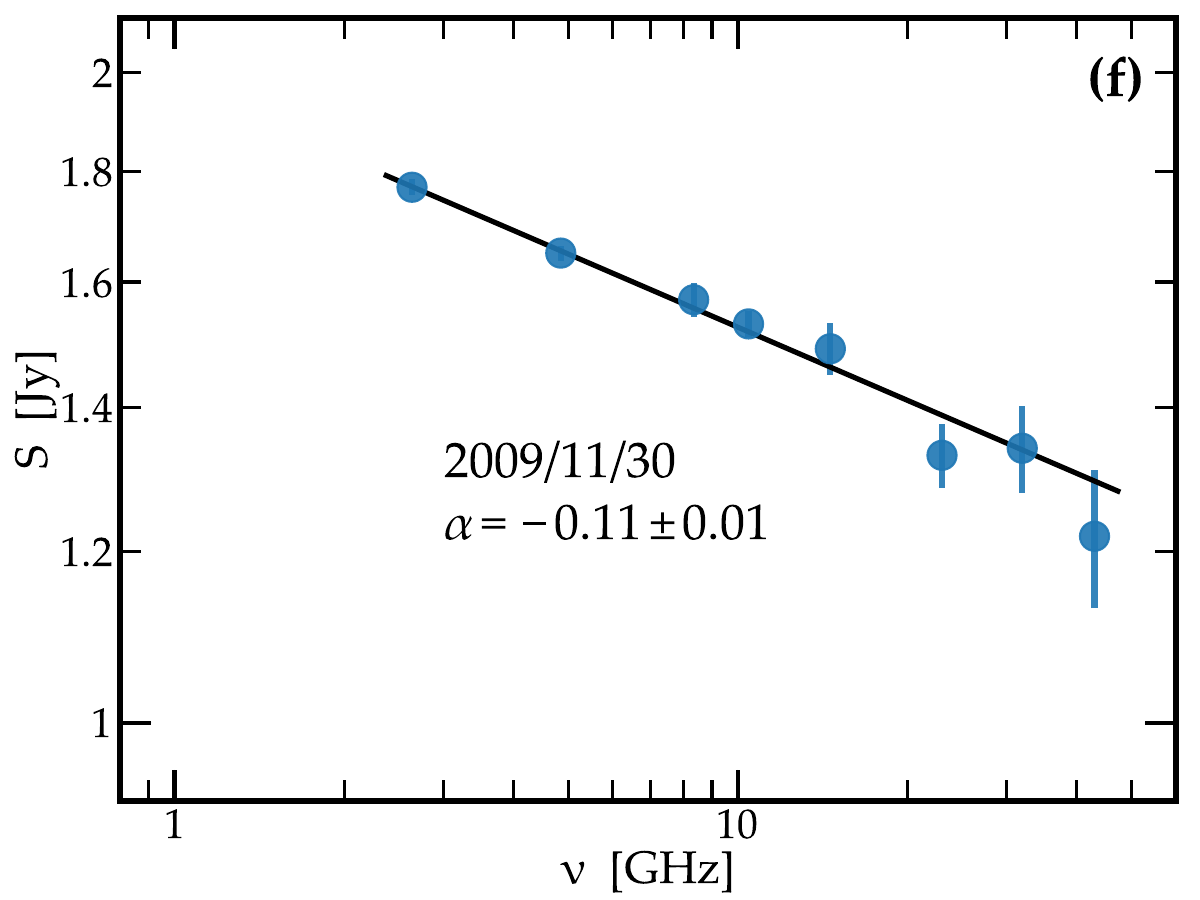}
\includegraphics[width=0.32\textwidth]{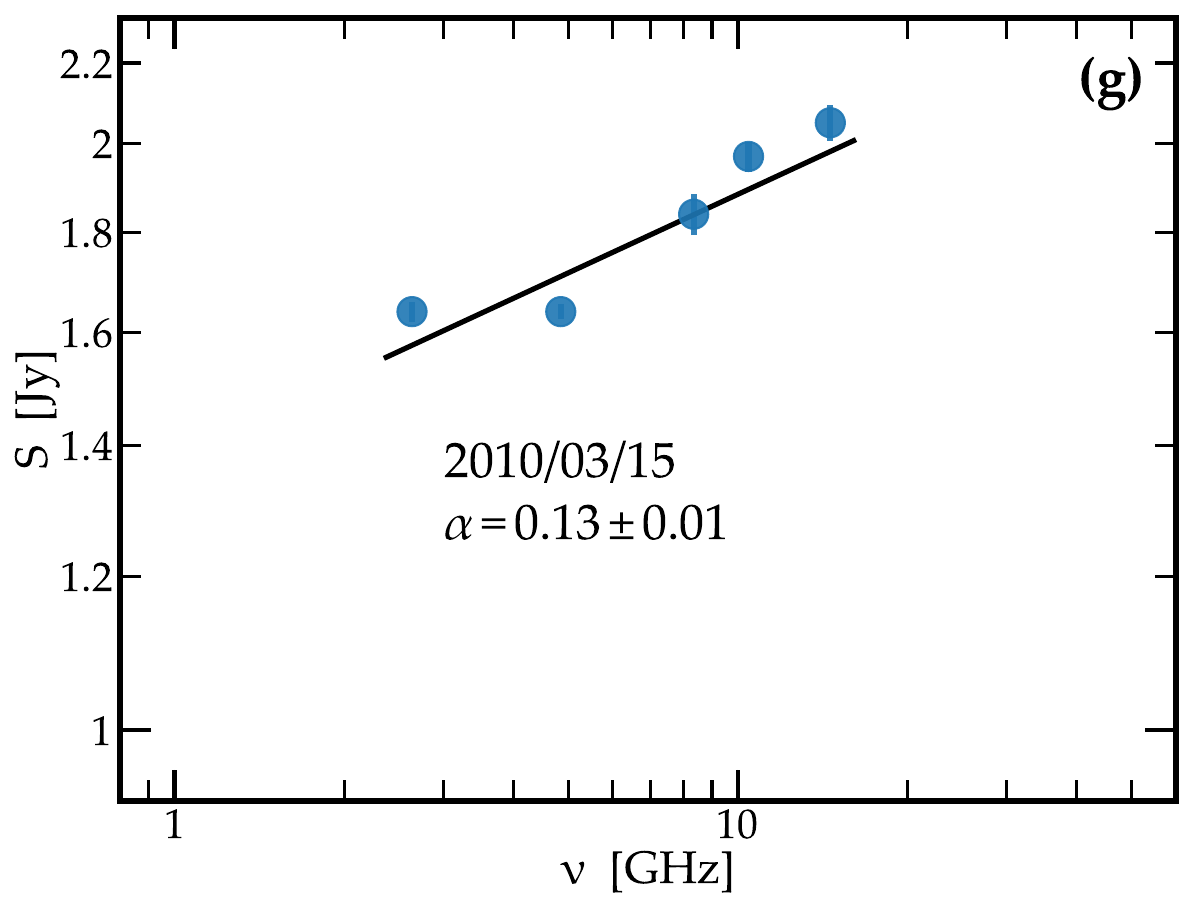}
\includegraphics[width=0.32\textwidth]{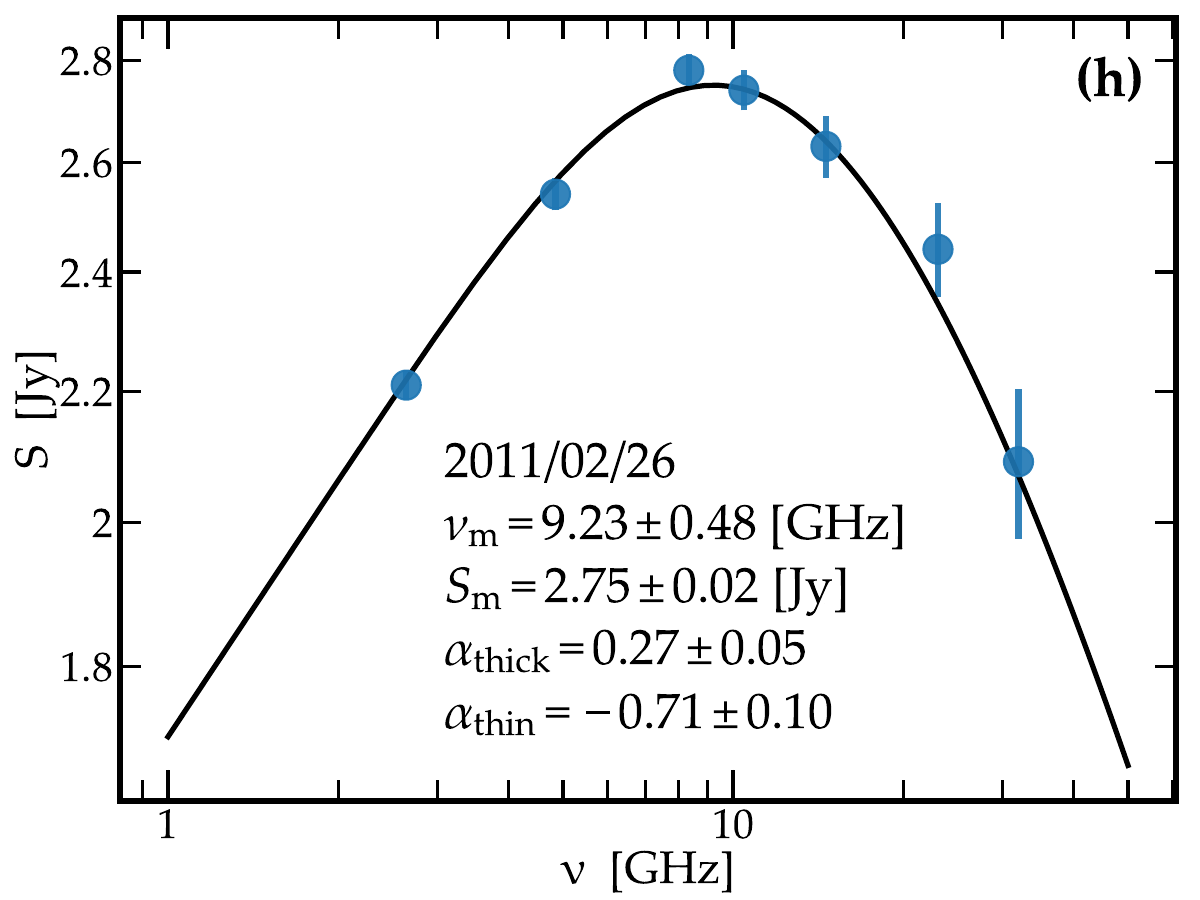}
\includegraphics[width=0.32\textwidth]{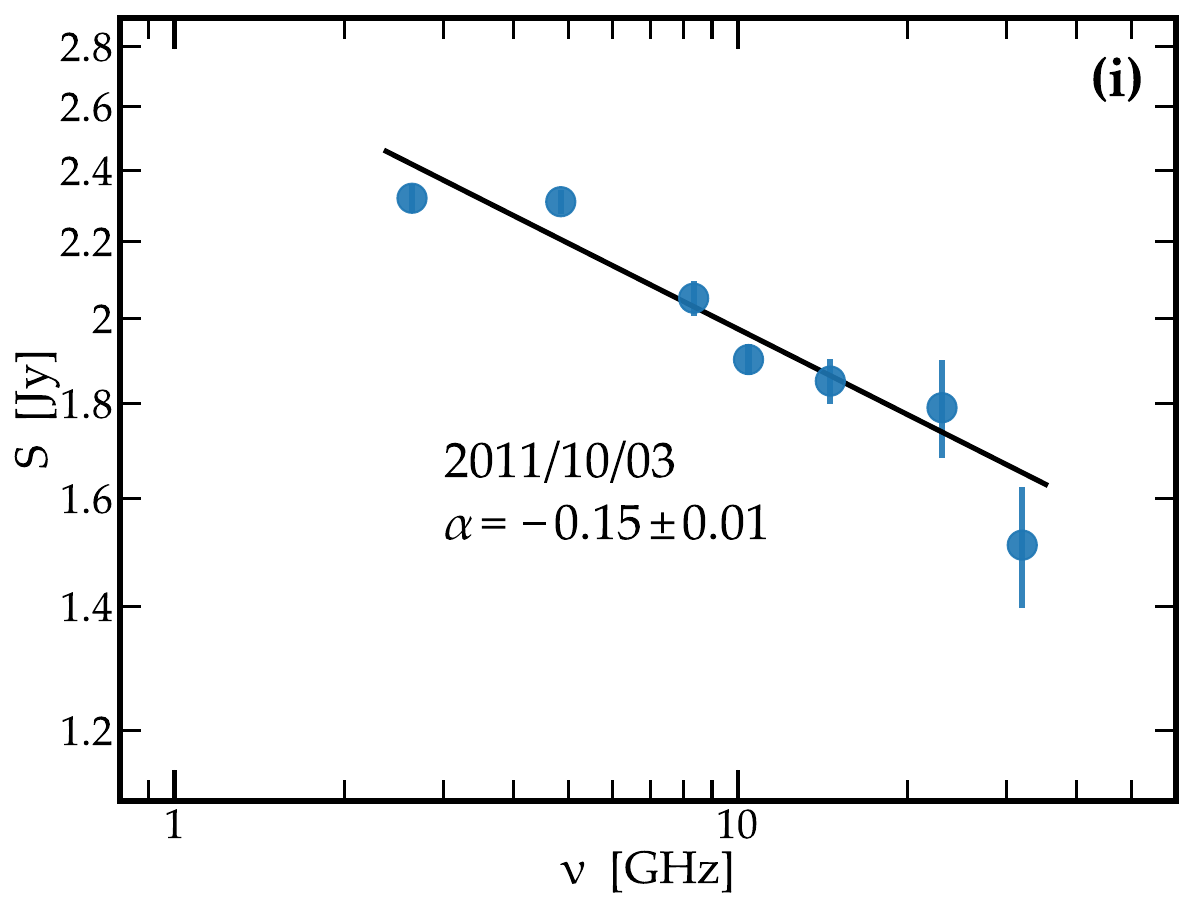}
\caption{Selected representative radio spectra during the 2007--2012 evolution}
\label{figure4}
\end{figure*}

The SSA spectra of 1156+295 exhibited multi-stage evolution during the five-year period from 2007 to 2012. By comparing with the radio flux variability (Figure~\ref{figure1}), it was found that most epochs exhibiting an SSA-peaked profile occur either near the radio flare events or during the subsequent stages with declining flux density. This is because the early increase of the radio flux density usually represents a transitional stage in which the radio emission arises from the tail of the previous shock, the onset of a new shock, or a superposition of both. Consequently, the radio spectrum within the observing bands (centimeter wavelengths) exhibits a monotonically decreasing (e.g., Figure~\ref{figure4}b), monotonically increasing (e.g., Figure~\ref{figure4}d), or concave shape (e.g., Figure~\ref{figure4}c), rather than a typical SSA-peaked spectrum. The turnover frequency likely lies outside the observed range, preventing a single SSA component from successfully fitting the flux density data. In such cases, we fit the spectrum with either a single or broken power law using weighted least squares in log space, and do not attempt SSA magnetic-field estimates. Selected representative examples of the radio spectra during the 2007--2012 evolution are shown in Figure~\ref{figure4}. During the three segments delineated in Section~\ref{section2.1}, the radio spectrum of 1156+295 exhibited the following evolution:
\begin{itemize}
\item \textbf{2007--2008:} Following the tail end of the previous radio flare, the radio spectra first exhibited SSA-peaked morphologies (Figure~\ref{figure4}a), but evolved into monotonically decreasing profiles in the centimeter band during the second half of 2007 through early 2008 (see Figure~\ref{figure4}b), indicating very low turnover frequencies (i.e., $\nu_{\rm m}<2.64$ GHz). This evolution suggests a continued decline in the turnover frequency.
\item \textbf{2008--2010:} During the rising stage of the radio flare (2008--2009), the spectrum evolved from monotonically decreasing to a concave shape (Figure~\ref{figure4}c) or a broad, plateau-like shape with an overall rising trend (Figure~\ref{figure4}d), suggesting the superposition of multiple SSA-peaked components. Following the flare in early 2009, the spectrum developed into a typical SSA-peaked shape (Figure~\ref{figure4}e). Throughout the subsequent flux density decay from 2009 to 2010, the SSA spectrum of 1156+295 underwent a continuous evolution in which both the turnover frequency $\nu_{\rm m}$ and peak flux density $S_{\rm m}$ generally showed a coordinated decline. By late 2009, the spectrum transitioned back to a monotonically decreasing profile (Figure~\ref{figure4}f).
\item \textbf{2010--2012:} Similarly to the evolution process observed between 2008 and 2010, but exhibiting a slightly more complex behavior during the fading stage, the radio spectrum of blazar 1156+295 evolved from a monotonically increasing spectrum (before the flare, Figure~\ref{figure4}g) to an SSA-peaked spectrum (after the flare, Figure~\ref{figure4}h), and eventually to a monotonically decreasing profile (Figure~\ref{figure4}i). This final monotonically decreasing morphology persisted until the next flare beyond 2012.
\end{itemize}

According to the shock-in-jet scenario \citep{1985ApJ...298..114M}, the radio spectral evolution reflects the various energy-accumulation/loss mechanisms within the jet, together with the emergence and expansion of possible new jet components.
The typical evolution can be parameterized by the relationship between the turnover flux density ($S_{\rm m}$) and turnover frequency ($\nu_{\rm m}$) as $S_{\rm m} \propto \nu_{\rm m}^{\epsilon}$, and can be broadly divided into three successive stages dominated by different energy-loss mechanisms \citep{2011A&A...531A..95F,2018ApJ...859..128A}: (i) the Compton stage, where $S_{\rm m}$ increases while $\nu_{\rm m}$ decreases, yielding an index $\epsilon < 0$ (e.g., $\epsilon = -2.5$ in \citealt{1985ApJ...298..114M}; $\epsilon = -0.4$ in \citealt{2000ApJ...533..787B}; and $\epsilon = -1.2$ in \citealt{2011A&A...531A..95F}); (ii) the synchrotron stage, where $S_{\rm m}$ remains roughly constant as $\nu_{\rm m}$ continues to decrease, yielding $\epsilon \approx 0$ \citep{1985ApJ...298..114M}; and (iii) the adiabatic stage, where both $S_{\rm m}$ and $\nu_{\rm m}$ decrease as the emitting region expands, yielding $\epsilon > 0$ (e.g., $\epsilon = 0.69$ in \citealt{1985ApJ...298..114M}; $\epsilon = 0.77$ in \citealt{2011A&A...531A..95F}).

\begin{figure*}[h!]
\centering
\includegraphics[width=0.47\textwidth]{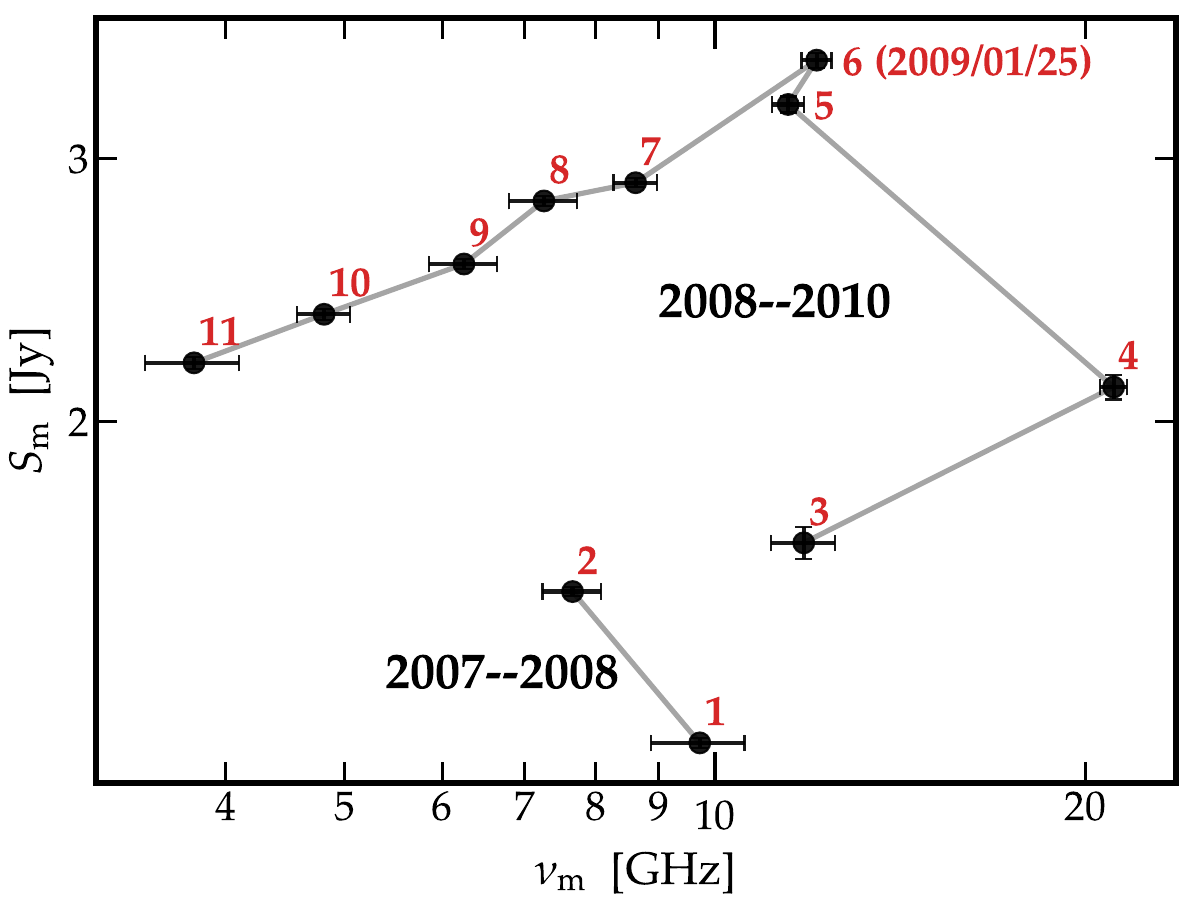}
\includegraphics[width=0.47\textwidth]{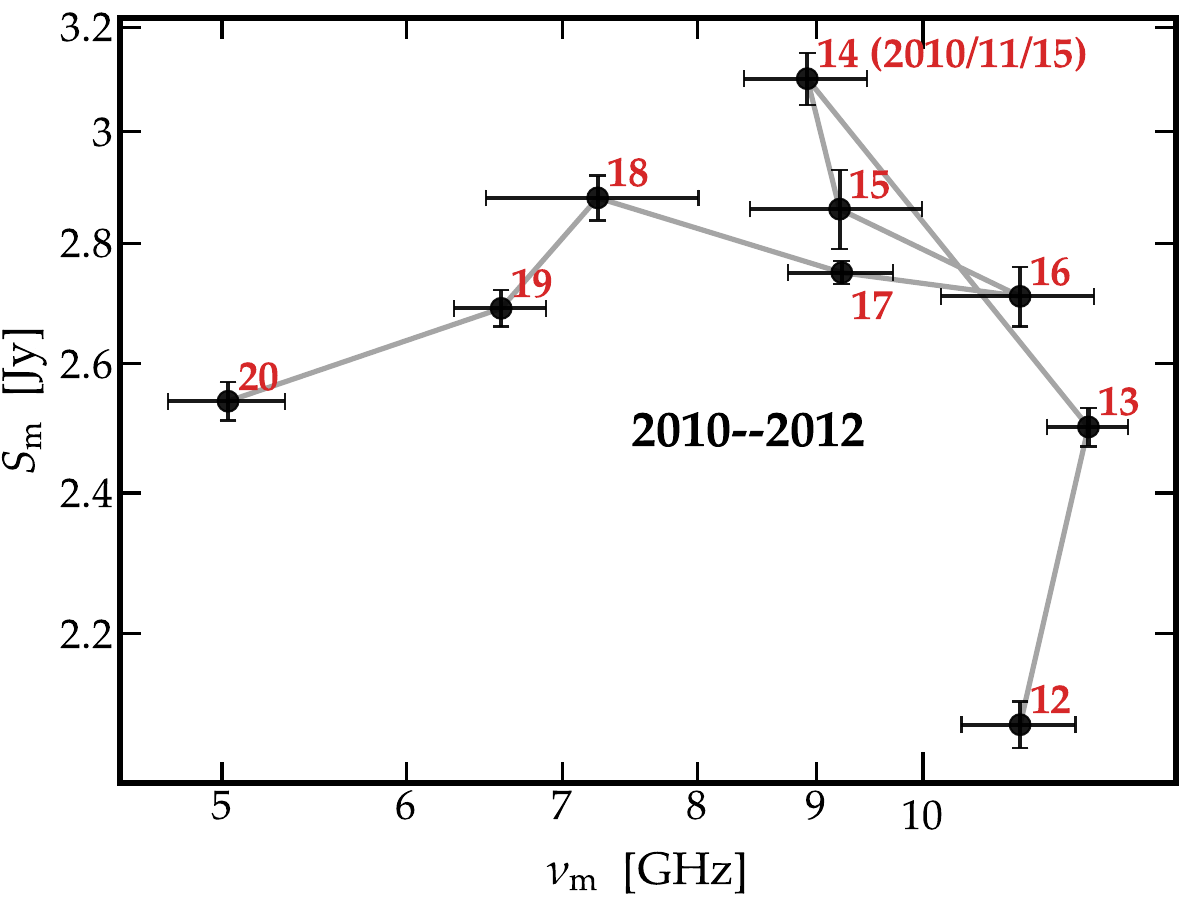}
\caption{Turnover flux density $S_{\rm m}$ vs. turnover frequency $\nu_{\rm m}$ for 1156+295 during 2009--2012.}
\label{figure5}
\end{figure*}

To investigate the spectral evolution of 1156+295 between 2007 and 2012, we plotted the turnover flux density against the turnover frequency in the $\nu_{\rm m}-S_{\rm m}$ plane, as shown in Figure \ref{figure5}. Between the two epochs in 2007--2008 (points 1--2 in Figure \ref{figure5}), $\nu_{\rm m}$ decreased while $S_{\rm m}$ increased, yielding a fitted index of $\epsilon = -0.98$, which corresponds to the Compton-dominated phase accompanied by energy injection associated with the onset of a new shock. Consistently, a new northward-directed jet component was indeed detected in the $Q$-band VLBI images in early 2008 \citep[e.g.,][]{2014MNRAS.445.1636R}. During the subsequent phase of rising flux density, the gap in the SSA spectral fitting results likely indicates a transitional energy-injection period, in which compression-driven particle acceleration and magnetic-field amplification cause both $\nu_{\rm m}$ and $S_{\rm m}$ to increase concurrently (points 2--4 in Figure \ref{figure5}). In this phase of simultaneous growth, the more pronounced variation in $\nu_{\rm m}$ suggests a possible $\nu_{\rm m}$ jump induced by the emergence of a new shock at high frequencies. Subsequently, the profound increase in $S_{\rm m}$ accompanied by a rapid decrease in $\nu_{\rm m}$ ($\epsilon = -0.79$ derived from points 4--6 in Figure \ref{figure5}) marks the main rising phase of the flare, which corresponds to the Compton-loss-dominated stage in the shock-in-jet framework. Following the radio flare in early 2009, both $\nu_{\rm m}$ and $S_{\rm m}$ clearly exhibited a simultaneous decline, yielding a derived index of $\epsilon = 0.40$ (left panel of Figure \ref{figure5}, points 6--11), which is consistent with the typical adiabatic expansion of the jet.

The SSA spectral fitting gap from late 2009 to early 2010 likewise corresponds to a transitional stage associated with the emergence of a new shock, during which the jet and its continuous evolution are detected in the VLBI images (see Figure~\ref{figure2}). The radio flux density and spectral evolution during 2010--2012 were broadly similar to those observed between 2008 and 2010, but with a more gradual flux density decline. Points 12--14 in Figure \ref{figure5} similarly correspond to an energy-injection phase, with the flare reaching its peak intensity around point 14 (2010/11/15). In the subsequent epochs (points 14--18), $S_{\rm m}$ remains nearly constant despite fluctuations in $\nu_{\rm m}$, indicating the synchrotron-loss-dominated stage. This feature may indicate a complex situation, in which jet expansion is accompanied by the emergence of a new inner jet component. The VLBI images support this inference, since multiple Gaussian components associated with the jet were detected in late 2010 \citep[e.g.,][]{2017ApJ...846...98J}. The successive emergence of multiple shocks slowed the flux density decline process and limited the conditions for further shock formation. As a result, after the late-2010 peak, the radio flux density remained in a slowly declining phase until a new major flare occurred beyond 2012 (around 2015; e.g., \citealt{2021A&A...651A..74K}). During this period, the evolution can essentially be classified as the adiabatic expansion stage, yielding an index of $\epsilon = 0.32$ extracted from points 18--20 in Figure~\ref{figure5}.
%It is not surprising that the Compton-loss stage in the shock-in-jet scenario is not observed, as this phase often drives the SSA turnover frequency above the typical centimeter observing bands \citep{1985ApJ...298..114M,2015A&A...580A..94F}.

\subsection{Magnetic Field Evolution}\label{section3.2}

\begin{figure*}[h!]
\centering
\includegraphics[width=1\textwidth]{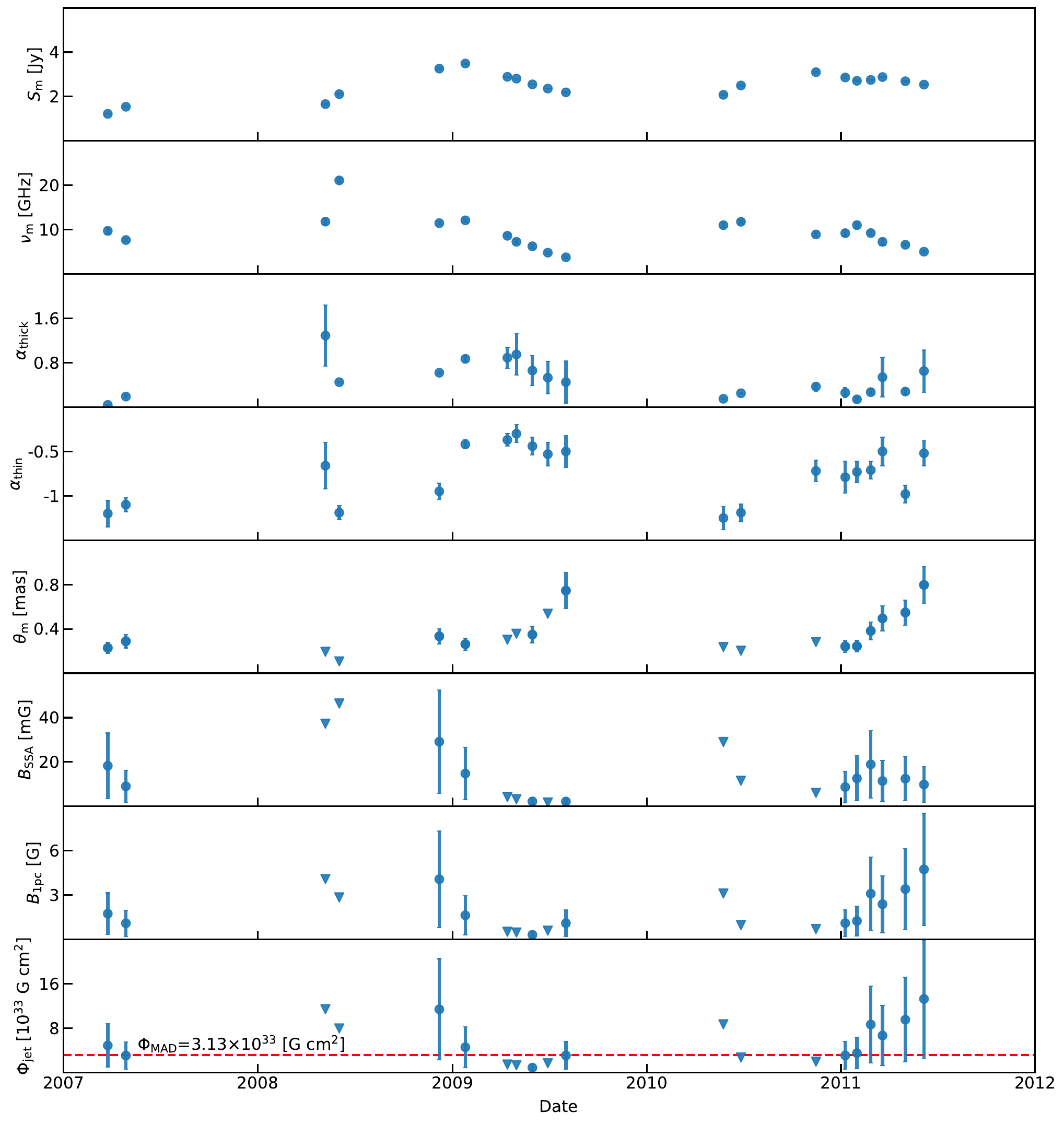}
\caption{Parameters and results of the magnetic field measurements. The downward-pointing triangles represent upper limits. The horizontal dashed line in the bottom panel indicates the predicted MAD magnetic flux of $\Phi_{\rm MAD}=3.13\times10^{33}$~G\,cm$^2$.}
\label{figure3}
\end{figure*}

The jet magnetic flux $\Phi_{\rm jet}$ is expressed in terms of the co-moving magnetic field strength in the jet at a distance of 1 pc from the central black hole $B_{\rm 1pc}$, as \citep{2015MNRAS.451..927Z}:
\begin{equation}
\Phi_{\rm jet} = 8 \times 10^{24} f(a_{*}) (1+\sigma)^{1/2} \left(\frac{M_{\rm BH}}{M_{\odot}}\right) \left(\frac{B_{\rm 1pc}}{\rm [G]}\right) ~ [\rm G~cm^2].
\label{equation9}
\end{equation}
Here, $f(a_{*}) = \frac{r_{\rm H}}{a_{*}r_g}=\frac{1+\sqrt{1-a_{*}^2}}{a_{*}}$ is a function of the black hole spin $a_{*}$. $r_{\rm H}=r_g(1+\sqrt{1-a_{*}^2})$ represents the black hole event horizon radius, and $r_g=GM/c^2$ is the gravitational radius of the black hole, with $G$ being the gravitational constant, $M$ the black hole mass, and $c$ the speed of light. To investigate the significance of magnetic flux in the Blandford--Znajek jet launching mechanism, we assume a maximally rotating central black hole ($a_{*}\sim1$), such that the spin-dependent efficiency factor $f(a_{*})\sim1$. In the equation above, $\sigma=\left( \Gamma \phi_{\rm j}/s \right)^2$ is the jet magnetization parameter, defined as the ratio of Poynting flux to kinetic-energy flux. Here, the parameter $s$ is a dimensionless factor of order unity that depends on the adopted magnetohydrodynamic scaling \citep{2009MNRAS.394.1182K}. For blazars, $\Gamma \theta_{\mathrm{j}} \sim 0.13$, which yields a value similar to those reported in the literature; additionally, $s=1$ is a common assumption \citep[e.g.,][]{2021A&A...652A..14C, 2024A&A...685L..11G}. For 1156+295, $M_{\mathrm{BH}}$ is the black hole mass of $10^{8.54}M_{\odot}$, as reported in the literature, which was estimated using the $\rm H\beta$ emission line \citep{1996ApJS..102....1B,2006ApJ...637..669L}. The parameters and results of the magnetic flux measurements are listed in Table~\ref{table1} and plotted in Figure~\ref{figure3}.

The MAD model proposes that relativistic jets are powered by strong magnetic fields which extract the rotational energy of the black hole \citep{1977MNRAS.179..433B}. When the accumulated magnetic flux threading the black hole approaches or exceeds a critical threshold, the release of magnetic energy efficiently drives the jet and enhances its power \citep{2003PASJ...55L..69N}. GRMHD simulations have shown that the MAD state is achieved when the magnetic flux reaches a saturation value given by \citep{2011MNRAS.418L..79T,2012MNRAS.423.3083M}:
\begin{equation}
    \Phi_{\rm MAD} = 50 \left(\dot{M}r_{\rm g}^2c\right)^{1/2} = 2.4\times 10^{25} \left( \frac{M_{\rm BH}}{M_{\odot}}\right) \left( \frac{L_{\rm acc}}{1.26\times10^{47}~\rm erg~ s^{-1}}\right)^{1/2}.
    \label{equation10}
\end{equation}
Here, a maximally spinning black hole is assumed with a radiative efficiency of 0.4 \citep[e.g.,][]{2014Natur.510..126Z,2021A&A...652A..14C}, and $\dot{M}$ denotes the mass accretion rate. For 1156+295, we adopt a black hole mass of  $M_{\rm BH}=10^{8.54}M_{\odot}$ \citep{1996ApJS..102....1B,2006ApJ...637..669L}. The accretion luminosity of the black hole is taken as $L_{\rm acc}=10^{46.25}~\rm erg~s^{-1}$, as estimated from the $\rm H{\beta}$ luminosity \citep{2006ApJ...637..669L,2014Natur.510..126Z}. Using these parameters, the predicted MAD magnetic flux of 1156+295 is $\Phi_{\rm MAD}=3.13\times10^{33}$~G\,cm$^2$, as shown in the bottom panel of Figure~\ref{figure3}. 

By comparing the estimated jet magnetic flux with the predicted MAD magnetic flux, it is possible to test whether a system has reached the MAD state and to evaluate the role of magnetic fields in launching and driving relativistic jets. For 1156+295, our estimates indicate that the inferred jet magnetic flux at a distance of 1 pc from the central black hole is close to or exceeds the critical MAD threshold in most epochs. This suggests that the magnetic fields play a dominant role in driving and powering the relativistic jet in 1156+295. Consistently with our results, \citet{2014Natur.510..126Z} also found 1156+295 to satisfy the MAD prediction, based on the magnetic flux $\Phi_{\rm jet}^{\rm CS}\approx 4.68\times 10^{33}$~G\,cm$^2$ derived under the equipartition assumption from core-shift measurements in a single epoch.

In the MAD scenario, the magnetic field strength and jet power are expected to evolve in a coupled manner, with the magnetic field intermittently exceeding the MAD threshold, accompanied by variations in jet power and the emergence of new jet components. Although the radio flux density is not a direct tracer of the total jet power, the SSA turnover flux density can serve as a proxy for the jet activity level and energy injection in the emitting region. Therefore, investigating the correlation between $B_{\rm SSA}$ and $S_{\rm m}$ provides a meaningful way to explore the coupled evolution of magnetic fields and jet power on parsec scales. Given that the temporal sampling of magnetic field measurements is relatively sparse and that some results represent upper limits, only a qualitative analysis is conducted here. During the 2008--2010 flaring episode, $B_{\rm SSA}$ reaches a local maximum around 2008 December, preceding a local maximum in $S_{\rm m}$ around 2009 February, suggesting that magnetic-field enhancement may occur prior to the radio flare maximum. In this process, the jet magnetic flux rapidly decreased from a high level exceeding the MAD threshold, with the release of magnetic energy leading to the radio flare and subsequent adiabatic expansion of the jet. This temporal ordering is consistent with a magnetically driven-scenario involving magnetic flux accumulation and release, although the sparse sampling precludes a robust quantitative constraint on the associated time lag. During the second radio flaring episode between 2010 and 2012, neither $S_{\rm m}$ nor $B_{\rm SSA}$ shows a clear monotonic trend, but instead both exhibit more complex variability. This behavior suggests that the spectral and magnetic field evolution during this period is influenced by the superposition of multiple jet components as well as structural evolution within the jet.

The ejection and apparent motion of the jet components may be temporally correlated with the radio flares \citep[e.g.,][]{2002A&A...394..851S} and magnetic energy release events \citep[e.g.,][]{2017A&A...597A..80H}. VLBI images reveal the emergence of new jet components in early 2008 \citep{2014MNRAS.445.1636R}, mid-2009 (see Figure~\ref{figure2}), and late 2010 \citep{2017ApJ...846...98J}, respectively. These epochs coincide with low magnetic-field states following the magnetic energy release, suggesting a magnetically driven shock-in-jet scenario in 1156+295. In this framework, magnetic energy is released first and injected into the shock, while a newly formed shock can be resolved as a distinct component in VLBI images only after it becomes optically thin at the observing frequencies and reaches a sufficiently high brightness contrast. Correspondingly, a brief decrease in the magnetic field strength and magnetic flux on 2011 March 20 preceded the subsequent emergence of a new jet component, which later formed a more complex superposition with a pre-existing shock. Overall, the combined evolution of the magnetic field, radio flux density, and jet structure supports a magnetically driven jet scenario, in which variations in jet activity are closely linked to the accumulation and release of magnetic flux near the central engine.

\section{Conclusions}\label{section4}

In this work, we combine multi-frequency single-dish and VLBI data to investigate the spectral properties and magnetic-field evolution of 1156+295 from 2007--2012. This enables us to place magnetic-field measurements, radio spectral evolution, inner-jet structural changes, and flux-density variability on a common timeline. This time-domain linkage allows us to relate the inferred $B_{\rm SSA}$ and $\Phi_{\rm jet}$ to the radio flares  and to assess their consistency with magnetically driven shock-in-jet scenarios. Our main conclusions are summarized as follows:
\begin{itemize}
\item Using multi-frequency Effelsberg single-dish radio flux densities, we model the SSA spectra of 1156+295 and derive the turnover frequency ($\nu_{\rm m}$) and peak flux density at the spectral turnover ($S_{\rm m}$) to estimate the jet magnetic field. Multi-epoch VLBI imaging is used to resolve the core--jet structure, measure jet kinematics, and extract core sizes and brightness temperatures. Combining the VLBI and SSA results, we further estimate the SSA magnetic field strength $B_{\rm SSA}$ and the jet magnetic flux $\Phi_{\rm jet}$.
\item The analysis of the radio light curve, SSA spectral morphologies, and the evolution trends of $\nu_{\rm m}$ and $S_{\rm m}$ suggests that the radio variability of 1156+295 is consistent with the shock-in-jet scenario. The observed radio spectra evolve through different characteristic forms during distinct variability phases, reflecting transitions between energy injection and dissipation processes in the jet.
\item The jet magnetic flux inferred from SSA modeling reaches or exceeds the MAD threshold in most epochs prior to the radio flares, suggesting that the jet of 1156+295 could be magnetically driven. A possible temporal sequence is observed, in which magnetic flux release appears to precede the radio flare peak, accompanied by the injection of magnetic energy into the relativistic jet. The detection of new VLBI jet components during low magnetic-field states further supports a magnetically dominated jet scenario.
\end{itemize}

The analysis of the coupled evolution between the magnetic field strength and jet power is limited by the sparse temporal sampling and the presence of upper limits in some epochs, restricting our results to a qualitative assessment of relative trends. Denser radio monitoring would provide stronger constraints on the time delays between magnetic energy release and radio flares. Our magnetic field estimates combine single-dish SSA spectral modeling with VLBI imaging and assume that the VLBI core dominates the SSA-emitting region. Future quasi-simultaneous multi-frequency VLBI observations will allow the SSA spectral properties of the radio core to be revealed directly on mas scales. The assumption of energy equipartition between particles and the magnetic field may deviate from reality during jet outbursts. This could affect the spectral index in Equation~\ref{equation2} and the intrinsic brightness temperature in Equation~\ref{equation3}, thus reducing the reliability of the magnetic-field strength estimates. 
In future studies, incorporating more complex multi-component SSA modeling, coupled with higher-cadence multi-band observations, may further help reveal overlapping emission components and clarify the underlying physical processes in complex transitional stages.
%In addition, applying multi-component SSA modeling during transitional stages within the shock-in-jet framework may help disentangle overlapping emission components and clarify the underlying physical processes.

\begin{acknowledgments}
We thank the anonymous referee for comments and suggestions that helped improve the paper.
This work was supported by the National Key R\&D Program of China (grant Nos. 2022SKA0120102 and 2024YFA1611500). W.-C. Xu acknowledges support from the China Scholarship Council (grant No. 202504910180). L.C. acknowledges support from the Tianshan Talent Training Program (grant No. 2023TSYCCX0099). T.A. acknowledges support from the Xinjiang Tianchi Talent Program. 
%S.F. thanks the Hungarian National Research, Development and Innovation Office (NKFIH) for support (grants OTKA K134213 and TKP2021-NKTA-64).
This work was partly supported by the Urumqi Nanshan Astronomy and Deep Space Exploration Observation and Research Station of Xinjiang (XJYWZ2303) and the Central Guidance for Local Science and Technology Development Fund (grant No. ZYYD2026JD01). 
This work used resources from the China SKA Regional Centre prototype. We acknowledge the use of archival calibrated VLBI data from the Astrogeo Center data base maintained by Leonid Petrov. This study made use of VLBA data from the VLBA-BU Blazar Monitoring Program (BEAM-ME and VLBA-BU-BLAZAR; \url{http://www.bu.edu/blazars/BEAM-ME.html}), funded by NASA through the Fermi Guest Investigator Program. This research has made use of data from the MOJAVE database that is maintained by the MOJAVE team \citep{2018ApJS..234...12L}. The VLBA is an instrument of the National Radio Astronomy Observatory. The National Radio Astronomy Observatory is a facility of the National Science Foundation operated by Associated Universities, Inc.

\end{acknowledgments}

\clearpage
\appendix

\section{VLBI flux densities}

\begingroup
\renewcommand{\arraystretch}{0.9}

\begin{table}[H]
\centering
\tablenum{A.1}
\caption{Multi-frequency VLBI flux densities of 1156+295 with an assumed 10\% uncertainty. \label{tableA1}}
%\tabletypesize{\footnotesize}
\footnotesize
\begin{tabular}{cc cc cc cc}
\toprule\toprule
\multicolumn{2}{c}{$S$-band (2.3\,GHz)} & \multicolumn{2}{c}{$X$-band (8.4\,GHz)} & \multicolumn{2}{c}{$Ku$-band (15.4\,GHz)} & \multicolumn{2}{c}{$Q$-band (43\,GHz)} \\
\cmidrule(lr){1-2} \cmidrule(lr){3-4} \cmidrule(lr){5-6} \cmidrule(lr){7-8}
Epoch & Flux Density & Epoch & Flux Density & Epoch & Flux Density & Epoch & Flux Density \\
(yyyy/mm/dd) & (Jy) & (yyyy/mm/dd) & (Jy) & (yyyy/mm/dd) & (Jy) & (yyyy/mm/dd) & (Jy) \\
\midrule
2007/01/11&0.48$\pm$0.05&2007/01/11&1.15$\pm$0.12&2007/02/05&1.47$\pm$0.15&2007/06/13&0.63$\pm$0.06\\
2007/12/05&0.63$\pm$0.06&2007/12/05&0.48$\pm$0.05&2007/07/03&0.80$\pm$0.08&2007/07/12&0.55$\pm$0.06\\
2008/04/02&0.56$\pm$0.06&2008/04/02&0.73$\pm$0.07&2008/05/01&1.67$\pm$0.17&2007/08/06&0.33$\pm$0.03\\
2009/05/13&1.60$\pm$0.16&2009/05/13&2.04$\pm$0.20&2008/08/06&2.52$\pm$0.25&2007/08/30&0.36$\pm$0.04\\
2011/08/02&1.60$\pm$0.16&2011/01/09&2.69$\pm$0.27&2009/01/07&3.43$\pm$0.34&2007/09/29&0.60$\pm$0.06\\
&&2011/06/19&1.92$\pm$0.19&2009/06/03&2.14$\pm$0.21&2007/11/01&0.65$\pm$0.07\\
&&2011/08/02&1.81$\pm$0.18&2009/12/10&1.30$\pm$0.13&2008/01/17&0.84$\pm$0.08\\
&&2011/12/07&1.57$\pm$0.16&2010/08/06&2.90$\pm$0.29&2008/02/28&0.96$\pm$0.10\\
&&&&2010/09/29&3.37$\pm$0.34&2008/06/12&2.05$\pm$0.21\\
&&&&2010/12/24&3.08$\pm$0.31&2008/07/06&2.12$\pm$0.21\\
&&&&2011/06/06&1.93$\pm$0.19&2008/08/15&2.20$\pm$0.22\\
&&&&2011/09/12&1.59$\pm$0.16&2008/09/10&2.42$\pm$0.24\\
&&&&&&2008/11/16&2.55$\pm$0.25\\
&&&&&&2008/12/21&2.00$\pm$0.20\\
&&&&&&2009/01/24&2.56$\pm$0.26\\
&&&&&&2009/02/22&2.02$\pm$0.20\\
&&&&&&2009/04/01&1.94$\pm$0.19\\
&&&&&&2009/05/30&1.30$\pm$0.13\\
&&&&&&2009/06/21&0.58$\pm$0.06\\
&&&&&&2009/07/26&0.62$\pm$0.06\\
&&&&&&2009/08/16&0.41$\pm$0.04\\
&&&&&&2009/09/16&0.80$\pm$0.08\\
&&&&&&2009/10/16&1.07$\pm$0.11\\
&&&&&&2009/11/28&1.05$\pm$0.10\\
&&&&&&2010/01/10&0.94$\pm$0.09\\
&&&&&&2010/02/10&1.37$\pm$0.14\\
&&&&&&2010/03/06&1.60$\pm$0.16\\
&&&&&&2010/04/07&2.20$\pm$0.22\\
&&&&&&2010/04/10&2.01$\pm$0.20\\
&&&&&&2010/04/15&2.26$\pm$0.23\\
&&&&&&2010/05/19&1.89$\pm$0.19\\
&&&&&&2010/06/14&1.15$\pm$0.11\\
&&&&&&2010/08/01&1.97$\pm$0.20\\
&&&&&&2010/08/21&2.12$\pm$0.21\\
&&&&&&2010/09/18&2.19$\pm$0.22\\
&&&&&&2010/10/24&2.36$\pm$0.24\\
&&&&&&2010/12/04&1.84$\pm$0.18\\
&&&&&&2011/01/02&2.15$\pm$0.22\\
&&&&&&2011/02/04&1.49$\pm$0.15\\
&&&&&&2011/03/01&1.58$\pm$0.16\\
&&&&&&2011/04/21&1.69$\pm$0.17\\
&&&&&&2011/05/22&1.40$\pm$0.14\\
&&&&&&2011/06/12&1.02$\pm$0.10\\
&&&&&&2011/07/21&1.03$\pm$0.10\\
&&&&&&2011/08/23&1.31$\pm$0.13\\
&&&&&&2011/09/16&0.83$\pm$0.08\\
&&&&&&2011/10/16&1.14$\pm$0.11\\
&&&&&&2011/12/02&0.82$\pm$0.08\\
\bottomrule
\end{tabular}
%\tablecomments{}
\end{table}
\endgroup

\clearpage
\section{VLBI Gaussian model-fit parameters}

\begingroup
\renewcommand{\arraystretch}{0.92}
\begin{deluxetable}{cccccccc}[h!]
\tablenum{B.1}
\tablecaption{Core parameters from 43 GHz VLBI images. \label{tableB1}}
\tabletypesize{\footnotesize}
\tablehead{
\colhead{Epoch} & \colhead{$S_{\nu}$} & \colhead{Beam Size} & \colhead{$\theta_{\nu}$} & \colhead{$T_{\rm b,obs}$} & \colhead{$\delta$} & \colhead{$\Gamma$} & \colhead{$\theta_{\rm view}$}\\ 
\colhead{(yyyy/mm/dd)} & \colhead{(Jy)} & \colhead{(mas$\times$mas)} & \colhead{(mas$\times$mas)} & \colhead{(K)} & \colhead{} & \colhead{} & \colhead{($^{\circ}$)}
} 
\startdata
2007/06/13&0.54$\times$0.22&0.08$\times$0.03&0.62$\pm$0.06&2.7$\pm$1.1&5.4$\pm$2.2&7.3$\pm$0.9&10.3$\pm$3.1\\
2008/06/12&0.38$\times$0.18&$<$0.08$\times$0.04&2.06$\pm$0.21&$>$8.5&$>$16.9&$>$9.9&$<$2.4\\
2008/12/21&0.48$\times$0.22&0.13$\times$0.06&1.99$\pm$0.20&2.8$\pm$1.2&5.7$\pm$2.3&7.2$\pm$0.8&9.9$\pm$3.2\\
2009/01/24&0.39$\times$0.22&0.12$\times$0.04&2.19$\pm$0.22&4.6$\pm$1.9&9.1$\pm$3.8&7.3$\pm$0.8&6.1$\pm$3.2\\
2009/04/01&0.40$\times$0.23&$<$0.08$\times$0.05&1.04$\pm$0.10&$>$3.3&$>$6.5&$>$7.1&$<$8.8\\
2009/05/30&0.38$\times$0.22&0.05$\times$0.05&0.72$\pm$0.07&3.2$\pm$1.3&6.4$\pm$2.6&7.1$\pm$0.5&8.9$\pm$3.3\\
2009/06/21&0.41$\times$0.21&$<$0.08$\times$0.04&0.39$\pm$0.04&$>$1.3&$>$2.5&$>$11.1&$<$14.5\\
2009/07/26&0.40$\times$0.23&0.08$\times$0.05&0.45$\pm$0.04&1.2$\pm$0.5&2.4$\pm$1.0&11.6$\pm$4.0&14.7$\pm$1.4\\
2010/05/19&0.41$\times$0.22&$<$0.08$\times$0.04&1.83$\pm$0.18&$>$5.8&$>$11.5&$>$7.9&$<$4.4\\
2010/06/14&0.39$\times$0.20&$<$0.08$\times$0.04&1.15$\pm$0.11&$>$4.2&$>$8.4&$>$7.2&$<$6.7\\
2010/12/04&0.39$\times$0.22&$<$0.08$\times$0.04&1.19$\pm$0.12&$>$4.0&$>$7.9&$>$7.1&$<$7.2\\
2011/01/02&0.42$\times$0.23&0.08$\times$0.03&1.74$\pm$0.17&7.4$\pm$3.0&14.7$\pm$6.1&9.0$\pm$2.3&3.0$\pm$2.0\\
2011/02/04&0.38$\times$0.22&0.11$\times$0.04&1.37$\pm$0.14&4.0$\pm$1.7&8.1$\pm$3.3&7.1$\pm$0.5&7.0$\pm$3.3\\
2011/03/01&0.47$\times$0.23&0.12$\times$0.06&1.47$\pm$0.15&2.5$\pm$1.0&5.0$\pm$2.1&7.5$\pm$1.2&10.8$\pm$3.0\\
2011/04/21&0.39$\times$0.22&0.08$\times$0.08&1.13$\pm$0.11&1.8$\pm$0.8&3.7$\pm$1.5&8.6$\pm$2.2&12.9$\pm$2.3\\
2011/06/12&0.41$\times$0.23&0.21$\times$0.04&1.02$\pm$0.10&1.4$\pm$0.6&2.7$\pm$1.1&10.5$\pm$3.4&14.2$\pm$1.6\\
\enddata
\tablecomments{$S_{\nu}$ is the flux density of the core component in the $Q$-band VLBI images. $\theta_{\nu}$ is the fitted core size or an upper limit taken as 1/5 of the restoring-beam major and minor axes. $T_{\rm b,obs}$ is the observed brightness temperature derived from $S_{\nu}$ and $\theta_{\nu}$, with its uncertainty estimated by adopting conservative systematic uncertainties of 10\% for the flux density and 20\% for the core size. The Doppler factor $\delta$ is estimated as the ratio of $T_{\rm b,obs}$ to an assumed intrinsic brightness temperature of 5$\times$10$^{10}$ K. The bulk Lorentz factor $\Gamma$ and the viewing angle $\theta_{\rm view}$ are calculated from $\delta$ and apparent proper motion $\beta_{\rm app}$. The derived parameters $T_{\rm b,obs}$, $\delta$, $\Gamma$ and $\theta_{\rm view}$ are subject to systematic uncertainties dominated by the assumed intrinsic brightness temperature and the core-size prescription. For epochs where the core is unresolved, $\theta_{\nu}$ is an upper limit, leading to lower limits on $T_{\rm b,obs}$ and $\delta$, and upper limits on $\theta_{\rm view}$. The lower limits on $\Gamma$ are obtained from Equation (\ref{equation5}) when $\delta_{\rm min}>\sqrt{\beta_{\rm app}^2+1} \approx 7.1$, otherwise we adopt the absolute minimum $\Gamma_{\rm min}=\sqrt{\beta_{\rm app}^2+1}$.}
\end{deluxetable}

\vspace{-2.5\baselineskip}

\begin{deluxetable}{cc}[h!]
\tablecaption{Core--jet component separations. \label{tableB2}}
\tablenum{B.2}
\setlength{\tabcolsep}{20.0pt}
\tabletypesize{\footnotesize}
\tablehead{
\colhead{Epoch} & \colhead{$R_{\rm core-jet}$} \\ 
\colhead{(yyyy/mm/dd)} & \colhead{(mas)}
} 
\startdata
2009/04/01&0.162$\pm$0.006\\
2009/05/30&0.165$\pm$0.001\\
2009/06/21&0.191$\pm$0.003\\
2009/07/26&0.236$\pm$0.004\\
2009/08/16&0.250$\pm$0.004\\
2009/09/16&0.301$\pm$0.002\\
2009/11/28&0.354$\pm$0.004\\
2010/01/10&0.396$\pm$0.006\\
2010/02/10&0.411$\pm$0.009\\
2010/03/06&0.413$\pm$0.004\\
2010/04/07&0.424$\pm$0.006\\
2010/04/10&0.426$\pm$0.004\\
\enddata
\end{deluxetable}
\endgroup

\clearpage
\section{Effelsberg Flux Density Data}

\begingroup
\renewcommand{\arraystretch}{0.95}
\begin{deluxetable}{ccccccccc}[h!]
\tablenum{C.1}
\tablecaption{Effelsberg single-dish flux density data used for the SSA spectral fitting. \label{tableC}}
\tabletypesize{\footnotesize}
\tablehead{
\colhead{Epoch} & \colhead{$S_{\rm 2.64GHz}$} & \colhead{$S_{\rm 4.85GHz}$} & \colhead{$S_{\rm 8.35GHz}$} & \colhead{$S_{\rm 10.45GHz}$} & \colhead{$S_{\rm 14.6GHz}$} & \colhead{$S_{\rm 23.05GHz}$} & \colhead{$S_{\rm 32GHz}$} & \colhead{$S_{\rm 43GHz}$}\\ 
\colhead{(yyyy/mm/dd)} & \colhead{(Jy)} & \colhead{(Jy)} & \colhead{(Jy)} & \colhead{(Jy)} & \colhead{(Jy)} & \colhead{(Jy)} & \colhead{(Jy)} & \colhead{(Jy)}
}
\startdata
2007/03/25*&1.21$\pm$0.02&1.13$\pm$0.01&1.25$\pm$0.02&1.29$\pm$0.02&1.14$\pm$0.03&&0.90$\pm$0.05&\\
2007/04/28*&1.32$\pm$0.01&1.48$\pm$0.02&1.53$\pm$0.02&1.49$\pm$0.02&1.43$\pm$0.03&1.04$\pm$0.03&0.79$\pm$0.03&0.86$\pm$0.07\\
2007/06/24&1.23$\pm$0.02&1.18$\pm$0.01&1.10$\pm$0.02&1.09$\pm$0.05&1.00$\pm$0.02&1.02$\pm$0.05&0.86$\pm$0.04&\\
2007/07/22&1.26$\pm$0.01&1.16$\pm$0.01&0.97$\pm$0.01&0.94$\pm$0.01&&0.74$\pm$0.08&0.70$\pm$0.04&\\
2007/08/19&&1.15$\pm$0.01&0.94$\pm$0.01&0.93$\pm$0.05&0.82$\pm$0.03&0.88$\pm$0.04&0.70$\pm$0.04&\\
2007/09/16&1.32$\pm$0.02&1.11$\pm$0.01&0.94$\pm$0.01&0.89$\pm$0.01&0.84$\pm$0.02&0.89$\pm$0.03&0.93$\pm$0.05&\\
2007/10/08&&0.98$\pm$0.01&0.85$\pm$0.01&0.78$\pm$0.01&0.77$\pm$0.02&0.94$\pm$0.05&0.92$\pm$0.03&0.99$\pm$0.05\\
2007/12/19&&0.95$\pm$0.01&0.89$\pm$0.01&0.88$\pm$0.01&0.86$\pm$0.02&0.85$\pm$0.04&&\\
2008/01/20&1.19$\pm$0.02&0.92$\pm$0.01&0.79$\pm$0.01&0.88$\pm$0.02&0.96$\pm$0.03&1.28$\pm$0.06&&\\
2008/02/18&1.17$\pm$0.02&0.87$\pm$0.01&0.83$\pm$0.01&0.93$\pm$0.01&1.18$\pm$0.03&1.45$\pm$0.05&&1.30$\pm$0.07\\
2008/05/06*&&1.05$\pm$0.01&1.51$\pm$0.02&1.68$\pm$0.02&1.76$\pm$0.07&1.71$\pm$0.18&1.69$\pm$0.13&\\
2008/06/01*&1.11$\pm$0.02&1.10$\pm$0.01&1.54$\pm$0.02&1.79$\pm$0.03&2.12$\pm$0.04&2.36$\pm$0.08&2.29$\pm$0.10&2.18$\pm$0.09\\
2008/09/19&&1.70$\pm$0.02&2.39$\pm$0.03&2.66$\pm$0.04&2.78$\pm$0.10&&3.83$\pm$0.12&2.23$\pm$0.13\\
2008/11/09&&2.13$\pm$0.02&2.94$\pm$0.05&3.12$\pm$0.06&3.00$\pm$0.06&3.29$\pm$0.13&&\\
2008/12/07*&1.63$\pm$0.01&2.37$\pm$0.02&3.06$\pm$0.04&3.27$\pm$0.05&3.25$\pm$0.08&3.15$\pm$0.18&&\\
2009/01/25*&1.78$\pm$0.02&2.71$\pm$0.02&3.37$\pm$0.05&3.47$\pm$0.06&3.49$\pm$0.09&3.33$\pm$0.17&3.06$\pm$0.10&3.00$\pm$0.11\\
2009/04/14*&2.00$\pm$0.03&2.65$\pm$0.02&2.89$\pm$0.03&2.88$\pm$0.04&2.80$\pm$0.06&2.53$\pm$0.10&2.29$\pm$0.08&2.30$\pm$0.16\\
2009/05/01*&2.28$\pm$0.03&2.72$\pm$0.02&2.83$\pm$0.04&2.79$\pm$0.04&2.64$\pm$0.06&2.35$\pm$0.13&2.35$\pm$0.08&\\
2009/05/31*&2.19$\pm$0.02&2.52$\pm$0.02&2.52$\pm$0.04&2.46$\pm$0.05&2.29$\pm$0.07&1.95$\pm$0.13&1.84$\pm$0.06&\\
2009/06/29*&2.19$\pm$0.02&2.36$\pm$0.02&2.25$\pm$0.03&2.10$\pm$0.03&1.98$\pm$0.05&1.61$\pm$0.13&1.43$\pm$0.13&\\
2009/08/02*&2.16$\pm$0.03&2.17$\pm$0.02&2.05$\pm$0.03&1.88$\pm$0.07&1.78$\pm$0.05&1.39$\pm$0.16&&\\
2009/08/30&2.09$\pm$0.02&1.95$\pm$0.02&1.76$\pm$0.03&1.67$\pm$0.02&1.46$\pm$0.06&1.18$\pm$0.07&&\\
2009/09/29&&1.80$\pm$0.02&1.67$\pm$0.02&1.57$\pm$0.02&1.48$\pm$0.04&1.14$\pm$0.04&0.79$\pm$0.03&\\
2009/11/05&1.84$\pm$0.02&1.66$\pm$0.01&1.56$\pm$0.03&1.49$\pm$0.02&1.42$\pm$0.03&&&\\
2009/11/30&1.77$\pm$0.02&1.65$\pm$0.01&1.57$\pm$0.03&1.53$\pm$0.02&1.49$\pm$0.04&1.33$\pm$0.05&1.34$\pm$0.06&1.22$\pm$0.09\\
2010/03/15&1.64$\pm$0.02&1.64$\pm$0.02&1.84$\pm$0.05&1.97$\pm$0.04&2.05$\pm$0.04&&&\\
2010/05/25*&1.75$\pm$0.02&1.87$\pm$0.02&2.08$\pm$0.03&&2.19$\pm$0.07&1.79$\pm$0.06&1.28$\pm$0.12&0.98$\pm$0.13\\
2010/06/27*&1.84$\pm$0.02&2.07$\pm$0.02&2.43$\pm$0.03&2.54$\pm$0.04&2.55$\pm$0.07&2.21$\pm$0.14&1.63$\pm$0.10&1.46$\pm$0.14\\
2010/08/15&&&2.88$\pm$0.04&3.02$\pm$0.22&2.86$\pm$0.10&1.73$\pm$0.10&2.23$\pm$0.14&\\
2010/10/17&2.25$\pm$0.02&2.80$\pm$0.03&3.24$\pm$0.08&3.34$\pm$0.08&3.46$\pm$0.14&&&\\
2010/11/15*&2.35$\pm$0.03&2.83$\pm$0.03&3.12$\pm$0.05&3.23$\pm$0.15&2.96$\pm$0.13&2.74$\pm$0.19&2.34$\pm$0.19&\\
2011/01/09*&2.30$\pm$0.03&2.65$\pm$0.03&&3.01$\pm$0.11&2.99$\pm$0.20&2.86$\pm$0.24&&\\
2011/01/31*&2.23$\pm$0.07&2.52$\pm$0.03&2.78$\pm$0.10&2.80$\pm$0.10&2.77$\pm$0.08&2.69$\pm$0.09&2.24$\pm$0.08&\\
2011/02/26*&2.21$\pm$0.02&2.54$\pm$0.03&2.78$\pm$0.03&2.74$\pm$0.04&2.63$\pm$0.06&2.44$\pm$0.08&2.09$\pm$0.11&\\
2011/03/20*&2.41$\pm$0.04&2.74$\pm$0.06&2.99$\pm$0.06&2.82$\pm$0.12&2.64$\pm$0.08&2.37$\pm$0.09&2.15$\pm$0.13&2.69$\pm$0.31\\
2011/05/02*&2.27$\pm$0.02&2.62$\pm$0.03&2.62$\pm$0.05&2.54$\pm$0.05&2.24$\pm$0.09&1.66$\pm$0.08&1.30$\pm$0.15&\\
2011/05/25&&2.58$\pm$0.03&2.49$\pm$0.03&2.38$\pm$0.04&2.30$\pm$0.06&2.01$\pm$0.08&1.87$\pm$0.14&\\
2011/06/06*&2.31$\pm$0.03&2.54$\pm$0.03&2.44$\pm$0.03&2.29$\pm$0.04&2.08$\pm$0.04&1.86$\pm$0.09&&\\
2011/07/10&2.39$\pm$0.02&2.49$\pm$0.03&2.31$\pm$0.03&2.19$\pm$0.04&1.99$\pm$0.05&1.80$\pm$0.12&1.26$\pm$0.09&\\
2011/08/07&2.39$\pm$0.03&2.40$\pm$0.03&2.19$\pm$0.05&2.04$\pm$0.11&1.88$\pm$0.15&&1.77$\pm$0.16&\\
2011/09/04&2.38$\pm$0.06&2.38$\pm$0.03&2.14$\pm$0.03&2.06$\pm$0.05&1.89$\pm$0.09&1.87$\pm$0.17&1.71$\pm$0.12&\\
2011/10/03&2.32$\pm$0.04&2.31$\pm$0.03&2.05$\pm$0.04&1.90$\pm$0.04&1.85$\pm$0.05&1.79$\pm$0.11&1.51$\pm$0.11&\\
2011/11/06&2.25$\pm$0.06&2.32$\pm$0.04&2.16$\pm$0.03&2.05$\pm$0.04&1.94$\pm$0.08&1.78$\pm$0.16&1.60$\pm$0.10&\\
2011/12/03&2.31$\pm$0.03&2.32$\pm$0.06&2.05$\pm$0.16&2.03$\pm$0.09&1.80$\pm$0.07&1.63$\pm$0.11&1.65$\pm$0.13&\\
\enddata
\tablecomments{SSA spectral fitting was restricted to epochs with at least five frequency measurements. Dates marked with an asterisk indicate epochs exhibiting an SSA-peaked spectrum; the corresponding plots are shown in Appendix~\ref{appendixD}.}
\end{deluxetable}
\endgroup

\clearpage
\section{Single-dish SSA Spectral Fitting Results}\label{appendixD}

\renewcommand{\thefigure}{D.1}
\begin{figure*}[htbp] 
\raggedright
\includegraphics[width=0.32\textwidth]{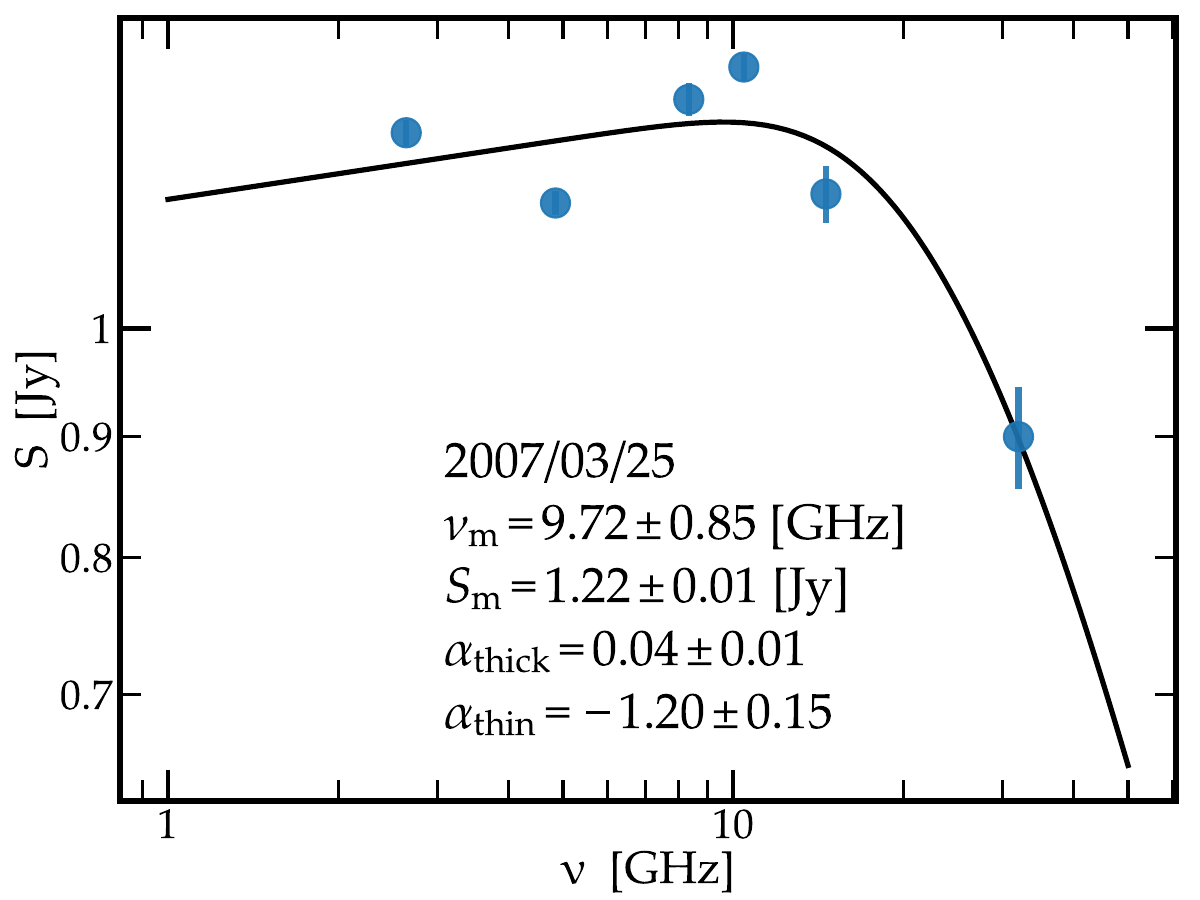}
\includegraphics[width=0.32\textwidth]{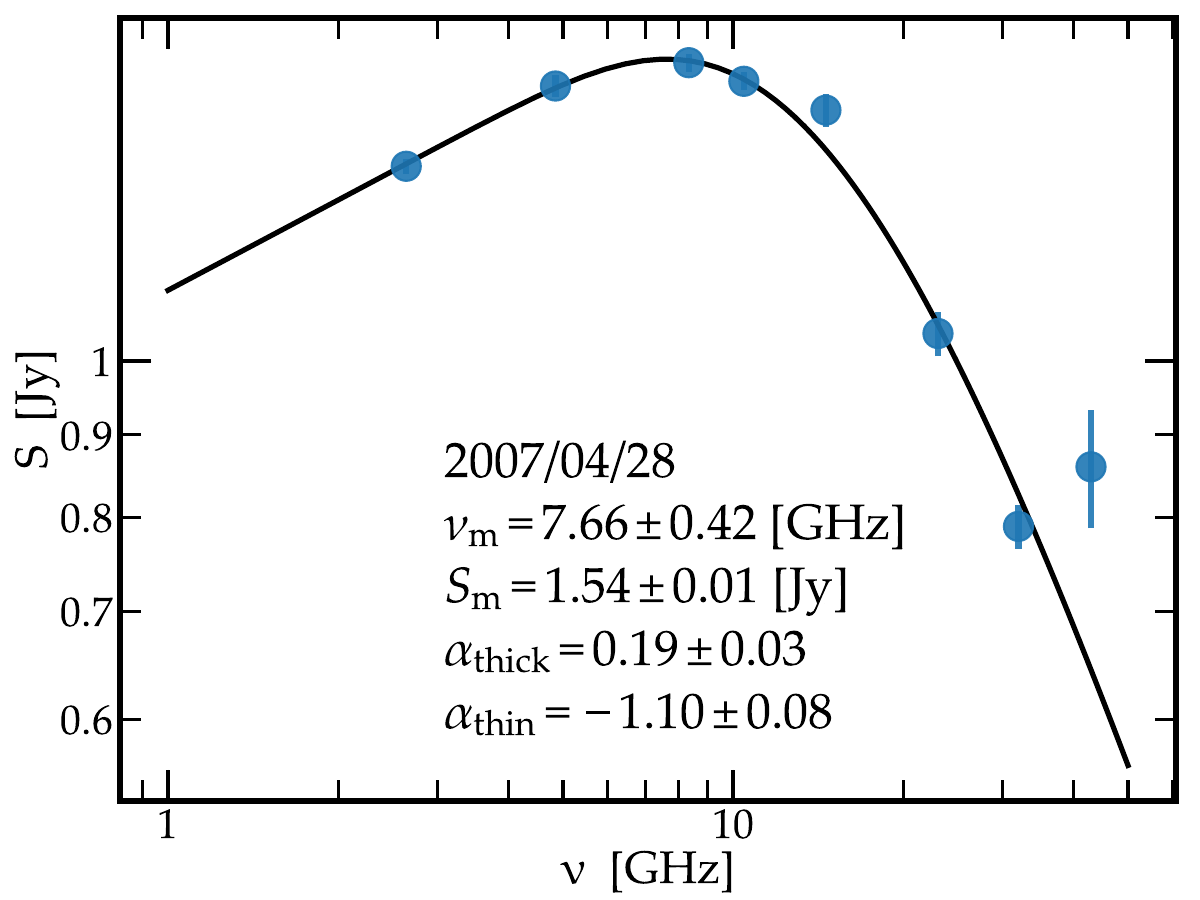}
\includegraphics[width=0.32\textwidth]{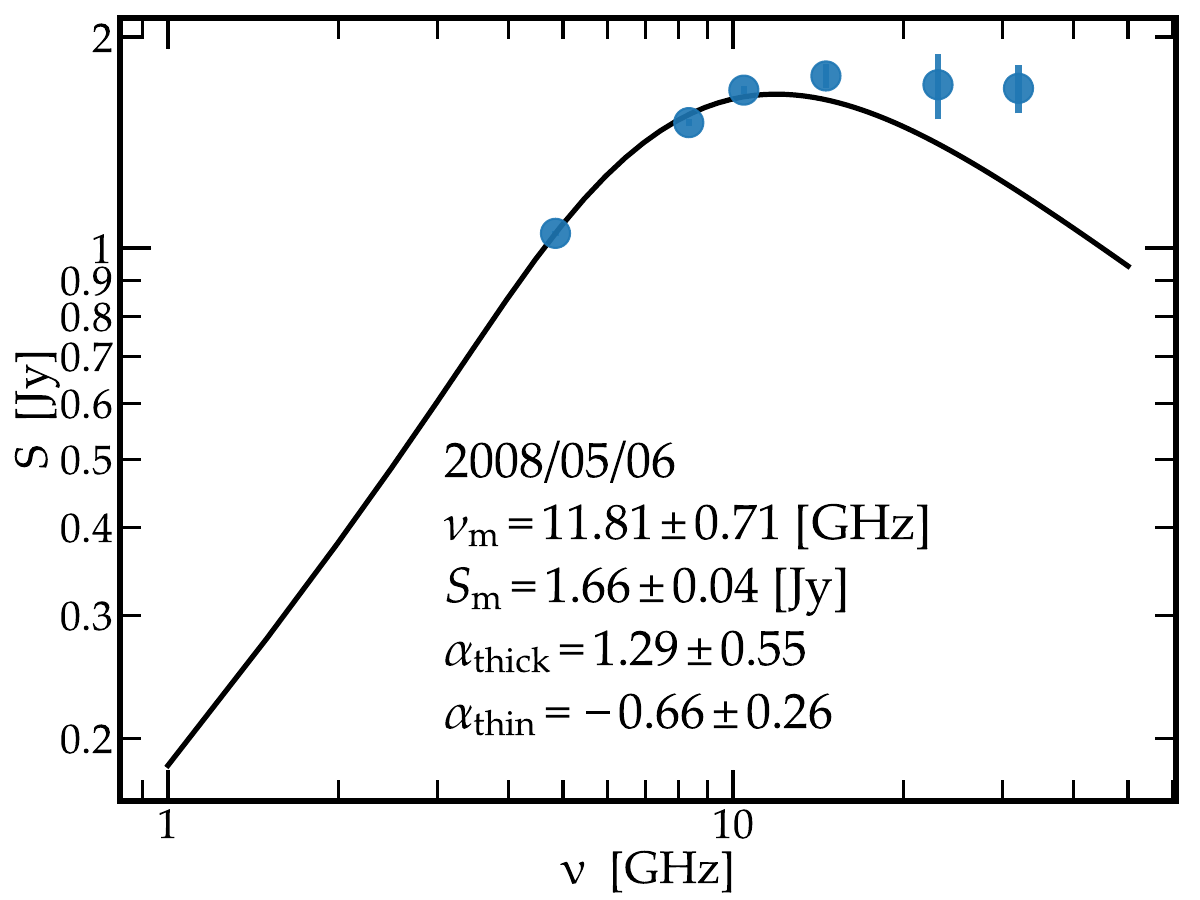}
\includegraphics[width=0.32\textwidth]{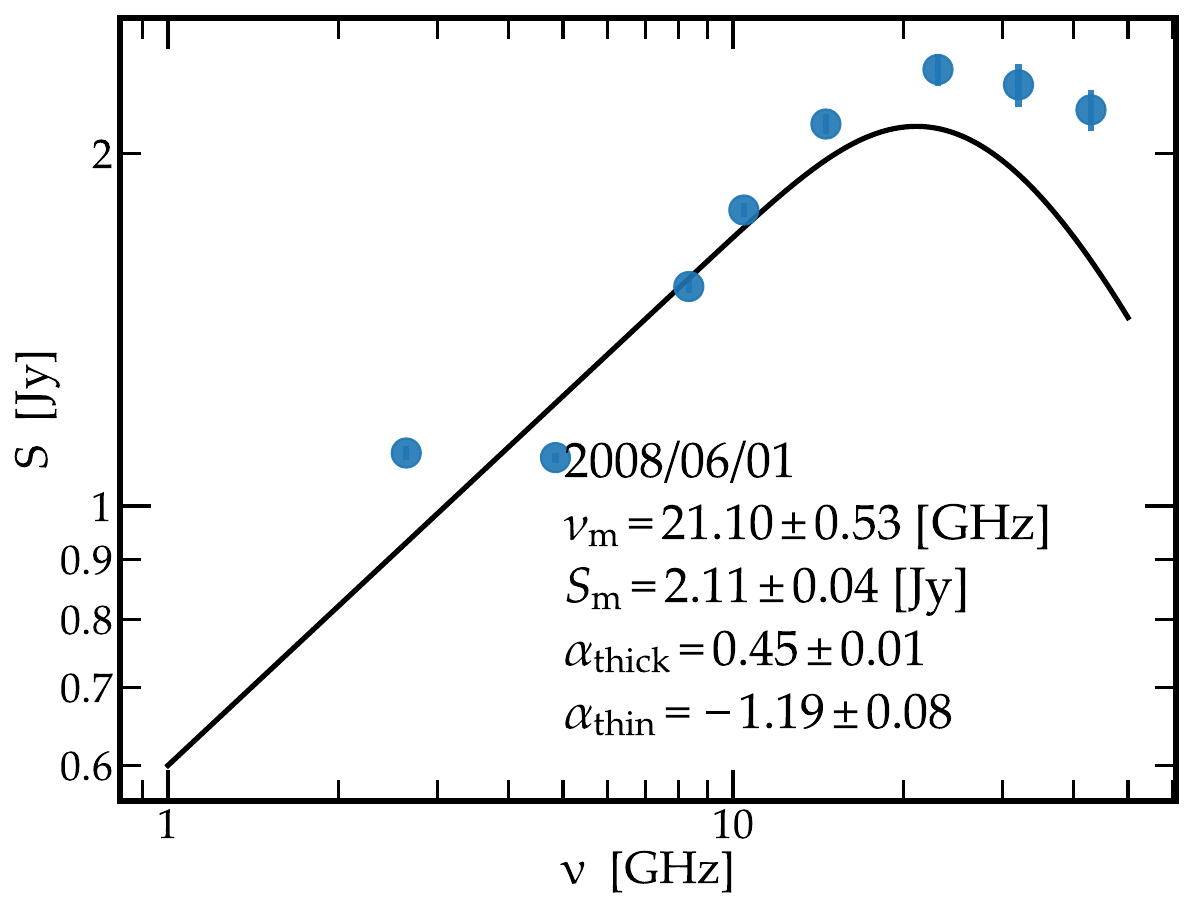}
\includegraphics[width=0.32\textwidth]{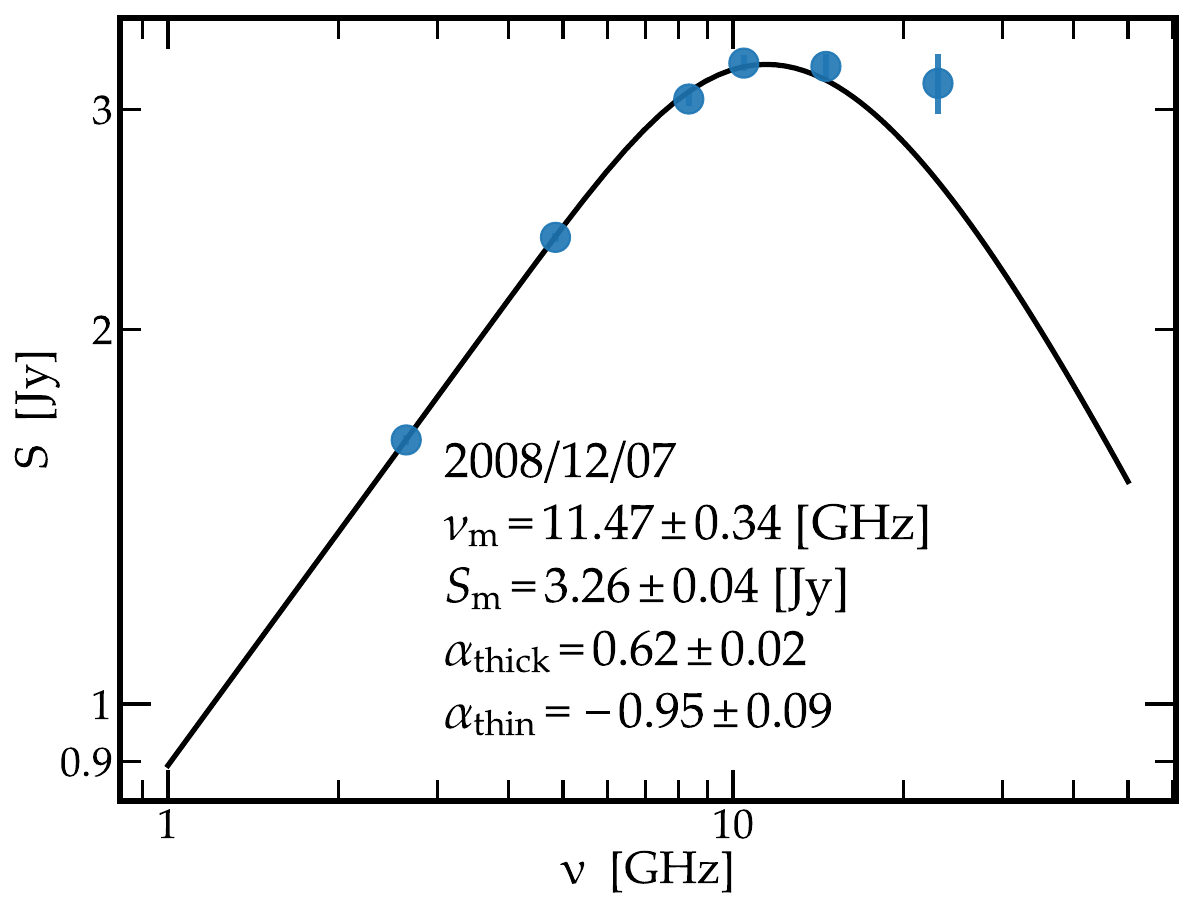}
\includegraphics[width=0.32\textwidth]{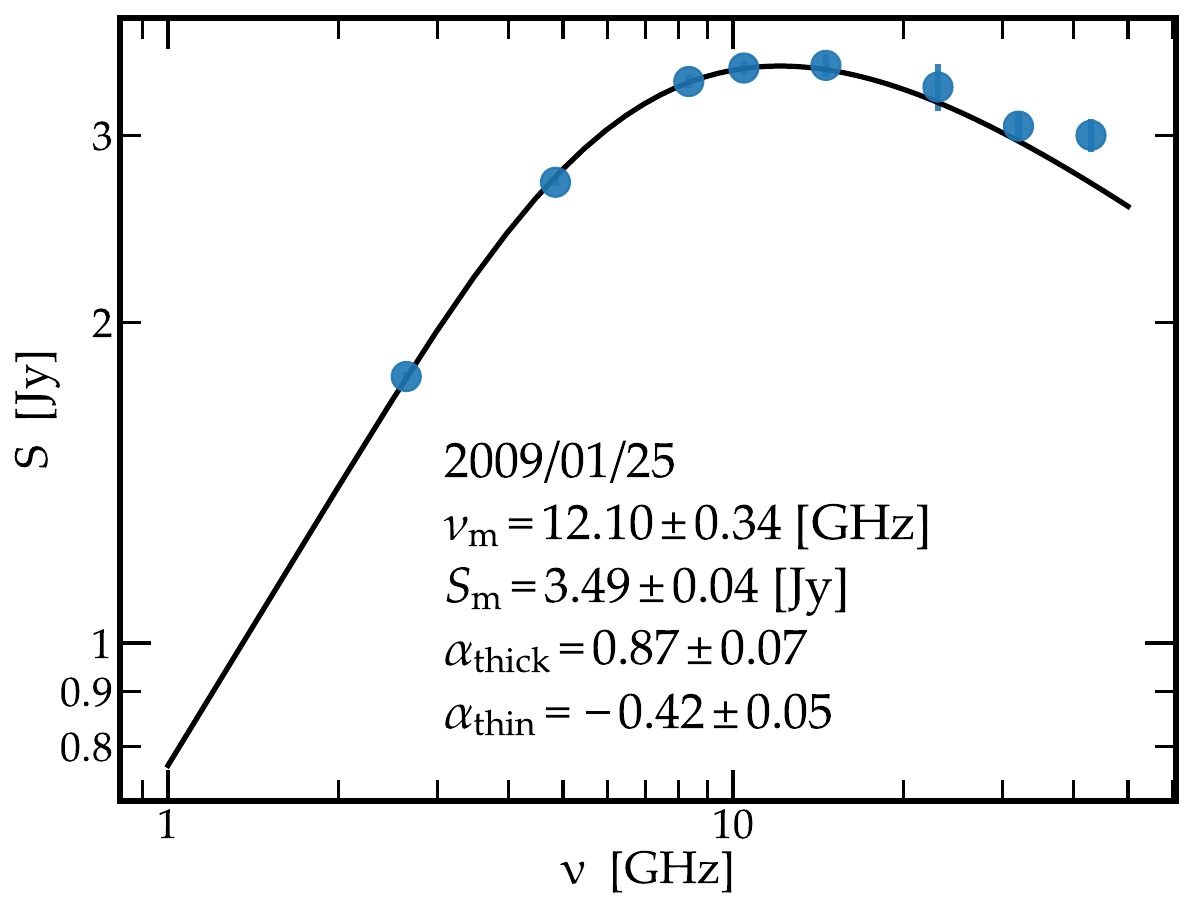}
\includegraphics[width=0.32\textwidth]{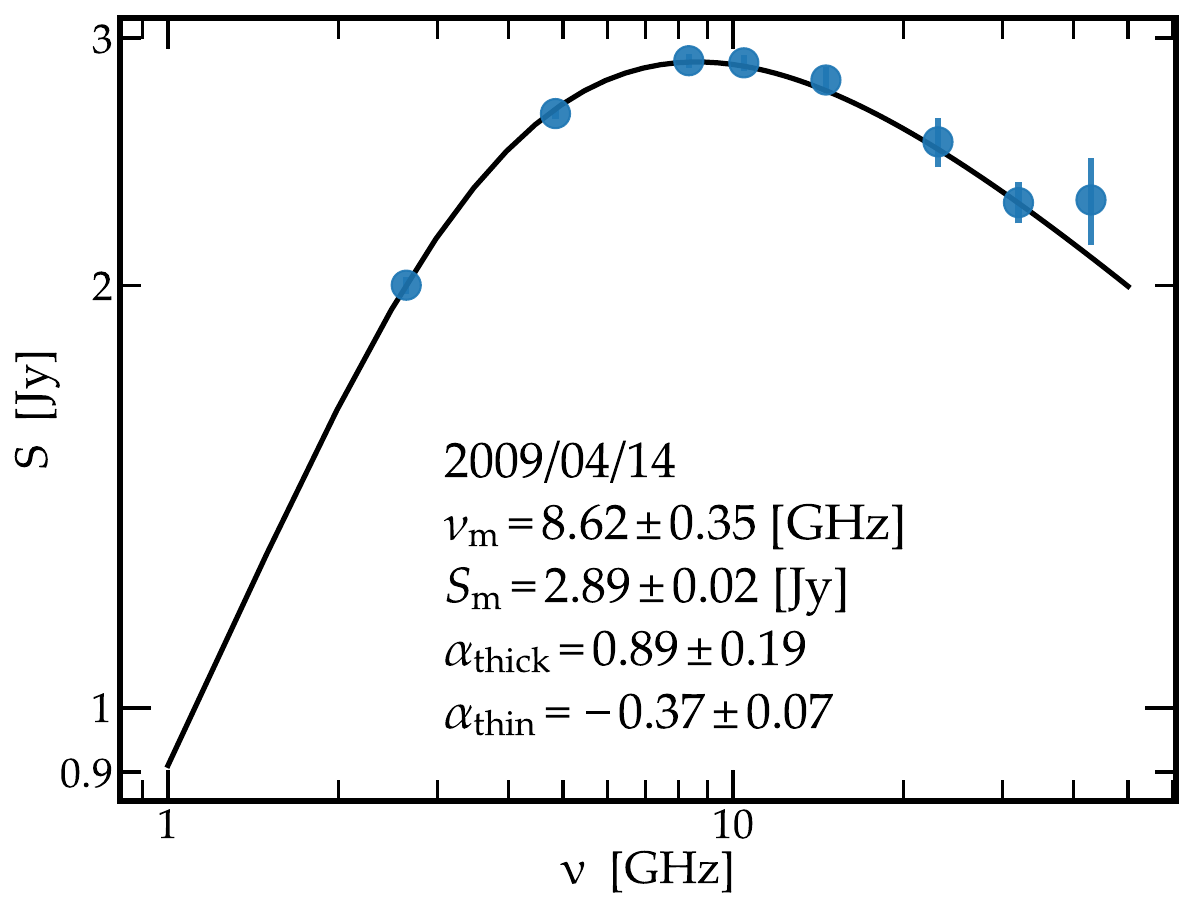}
\includegraphics[width=0.32\textwidth]{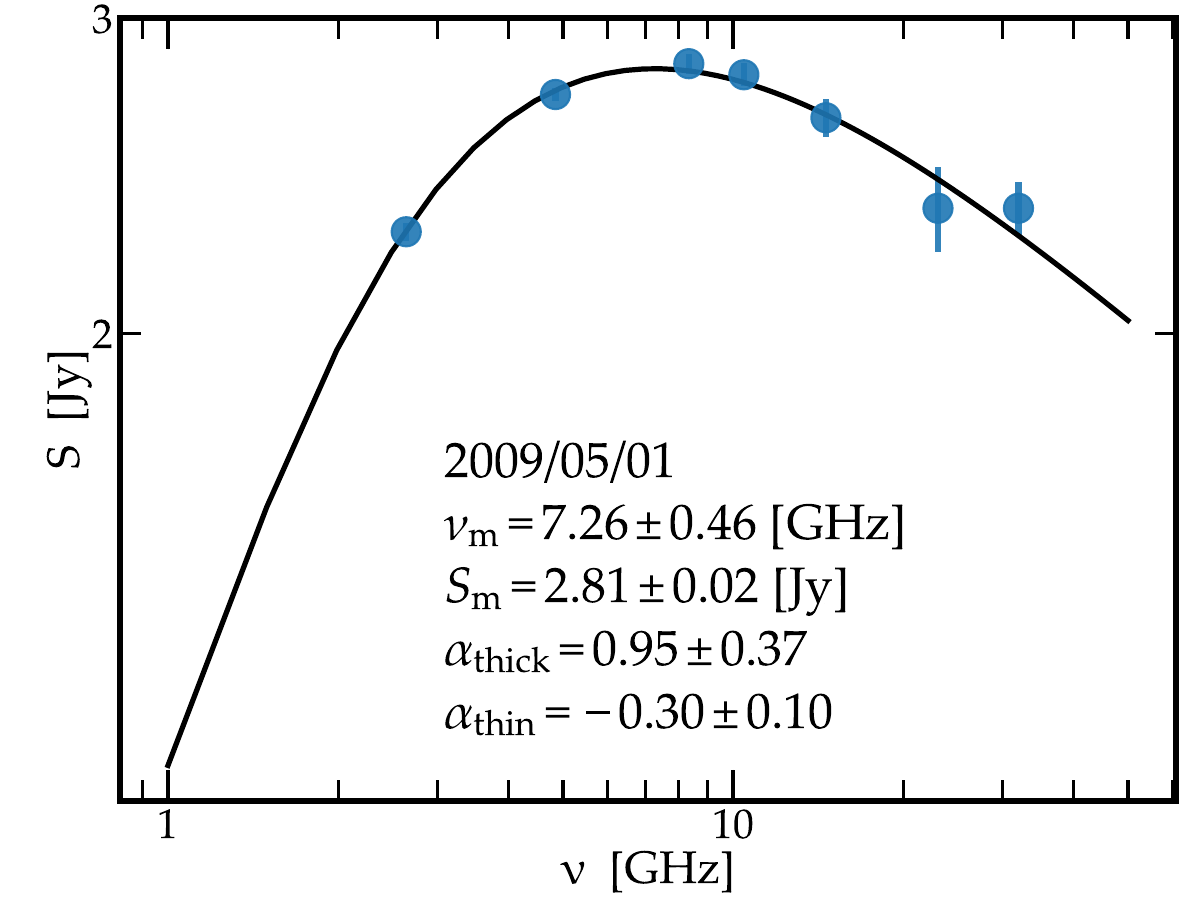}
\includegraphics[width=0.32\textwidth]{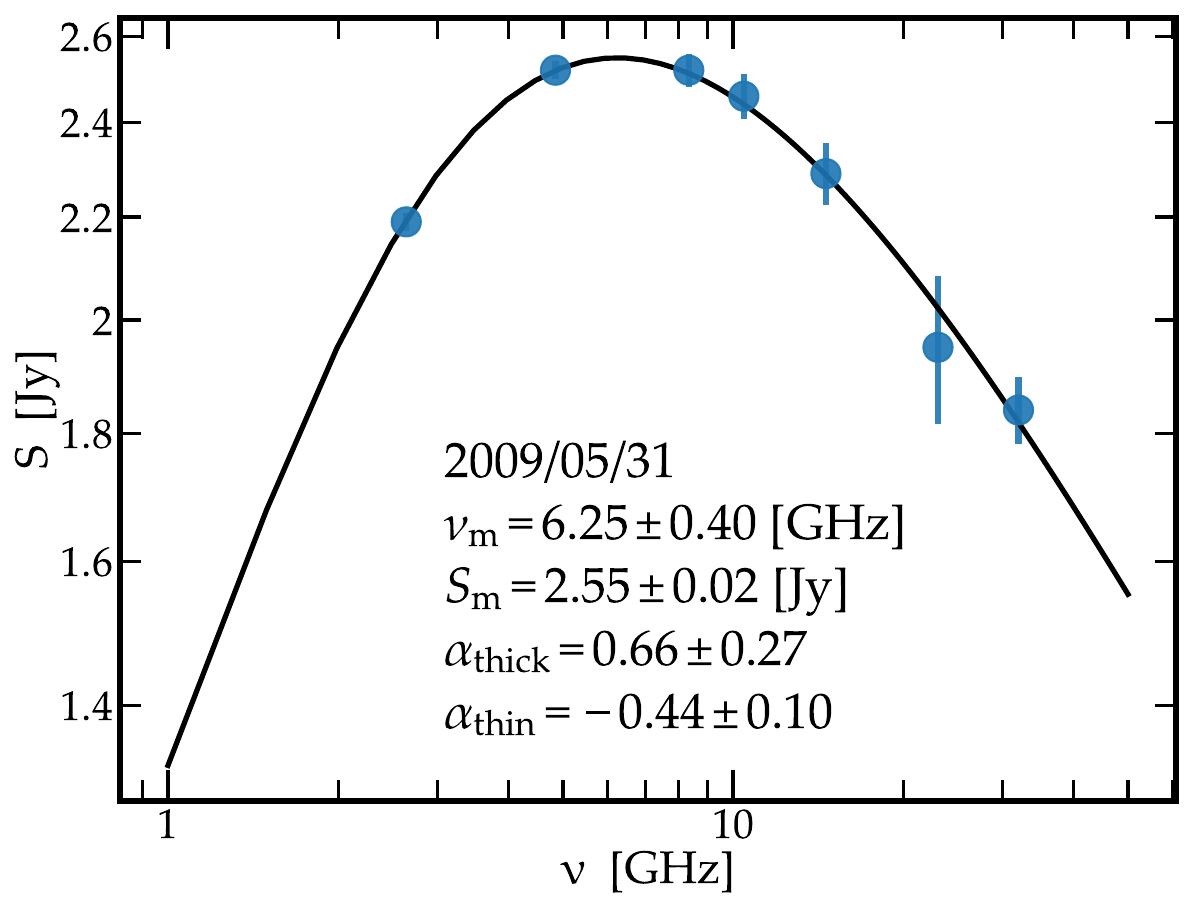}
\includegraphics[width=0.32\textwidth]{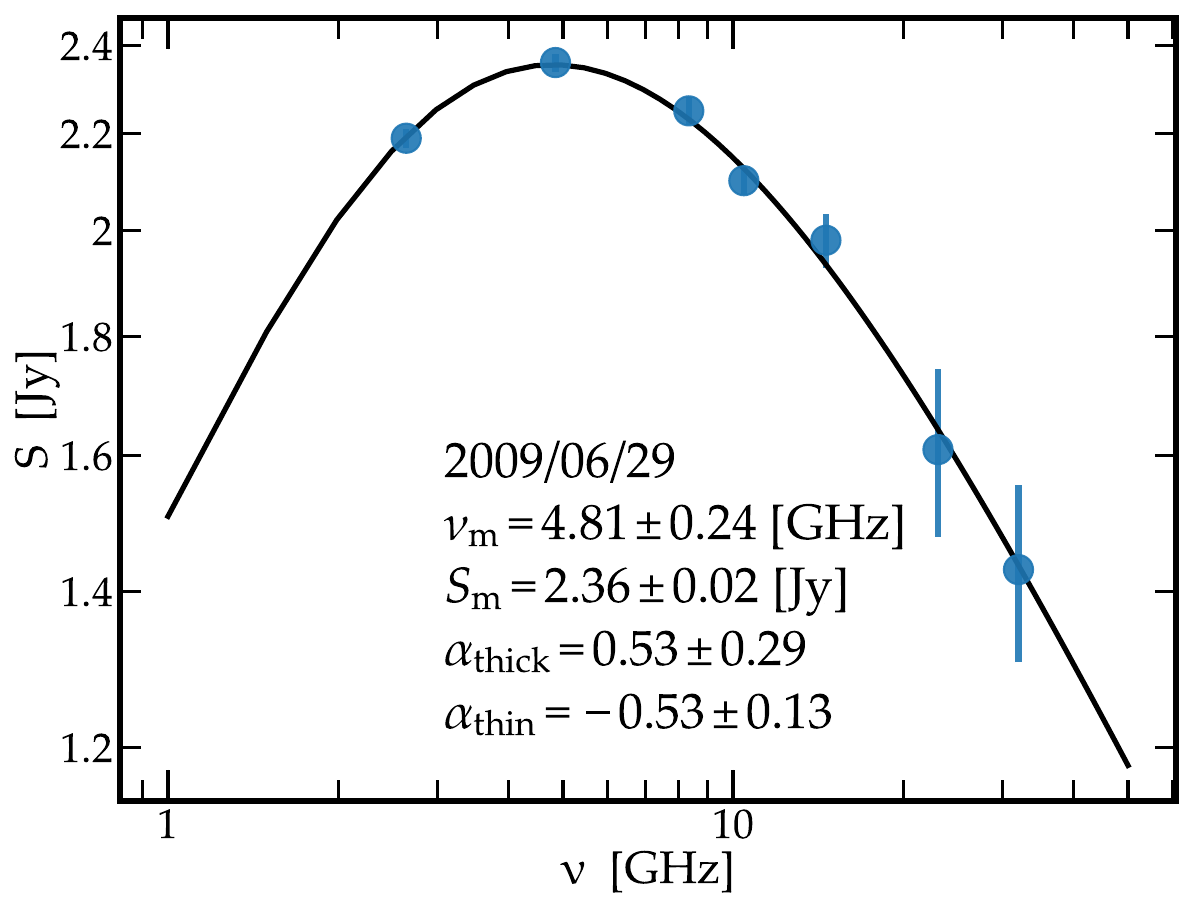}
\includegraphics[width=0.32\textwidth]{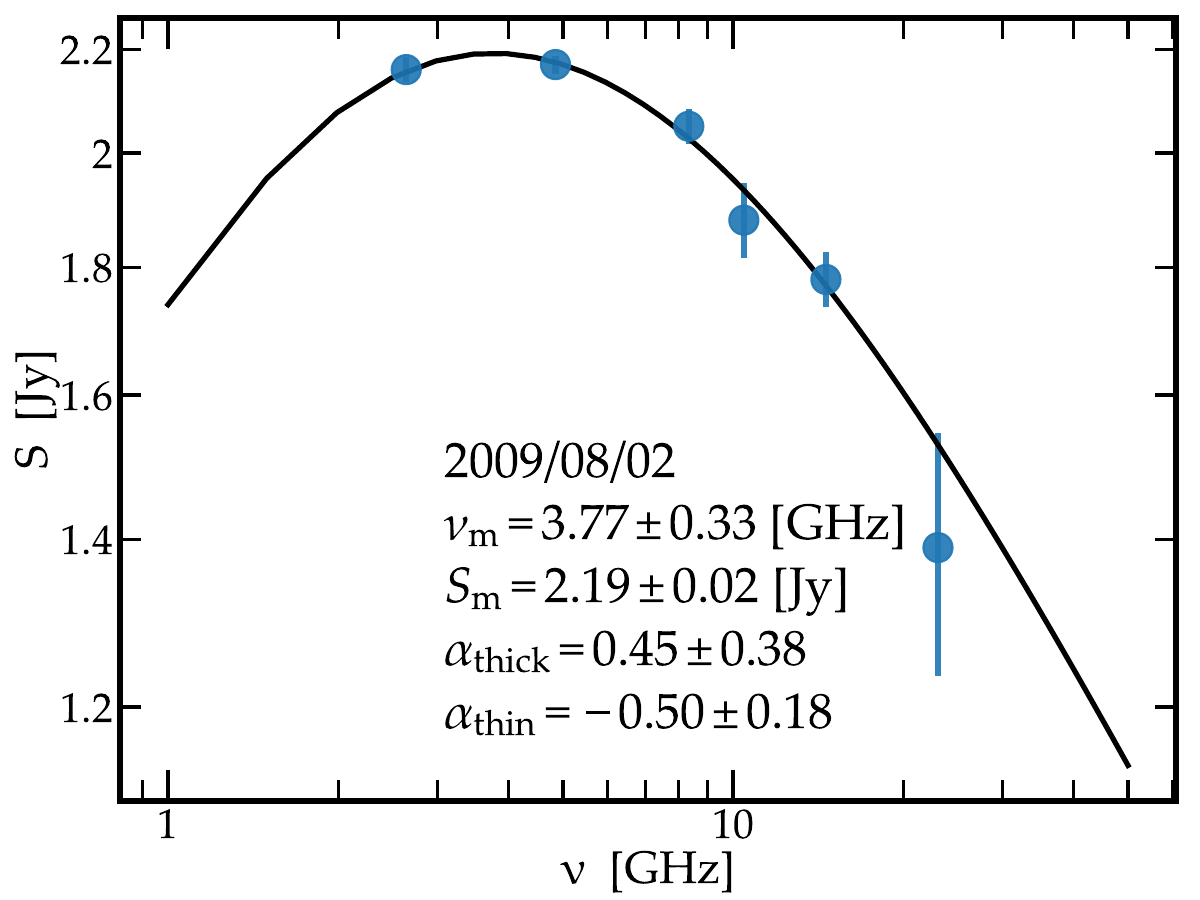}
\includegraphics[width=0.32\textwidth]{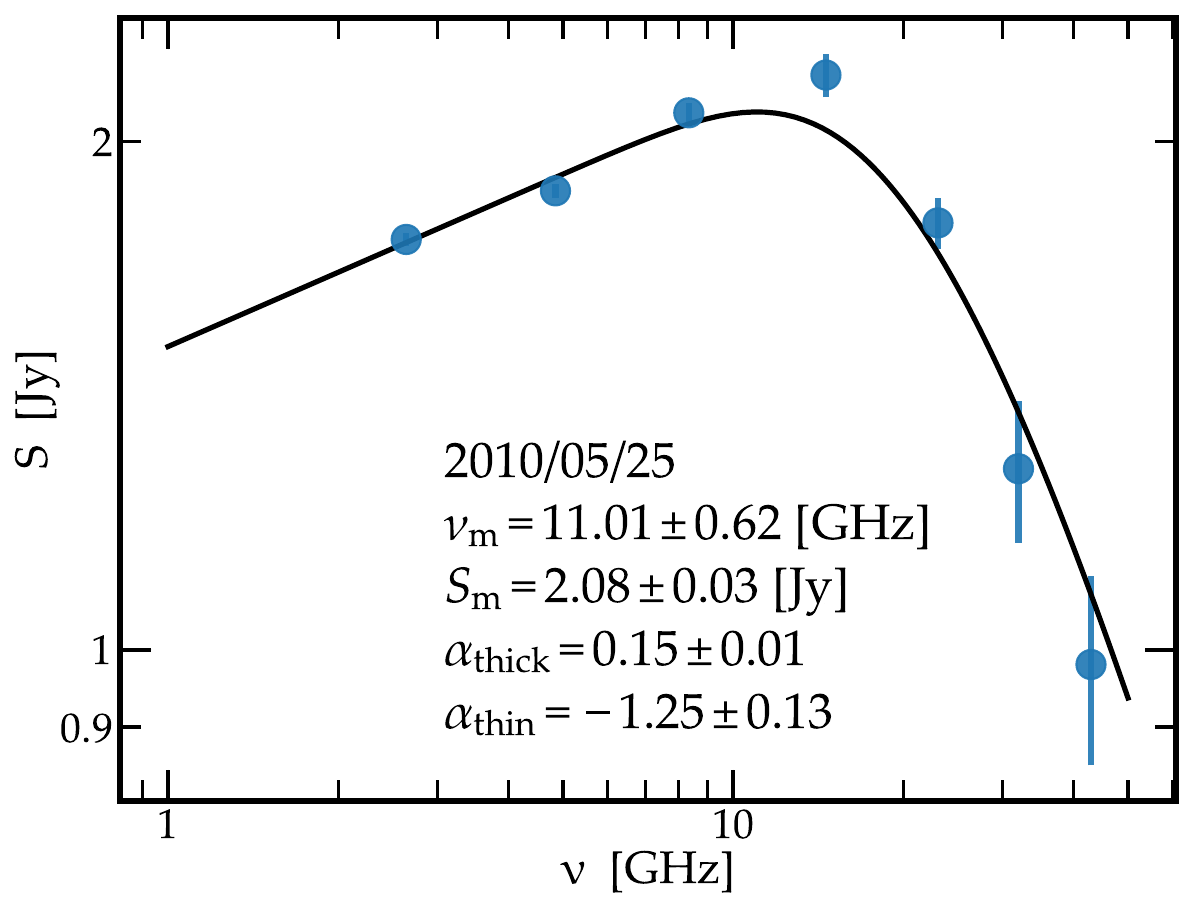}
\includegraphics[width=0.32\textwidth]{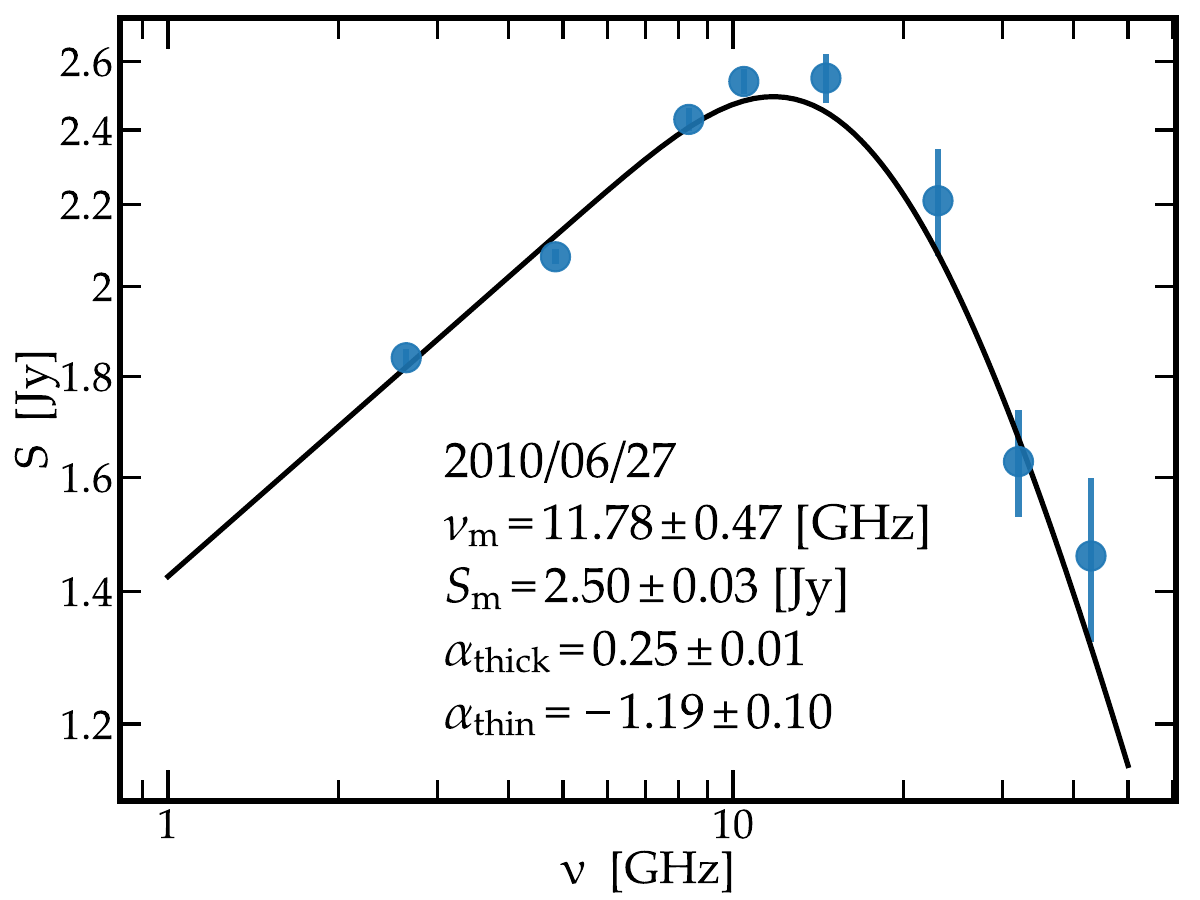}
\includegraphics[width=0.32\textwidth]{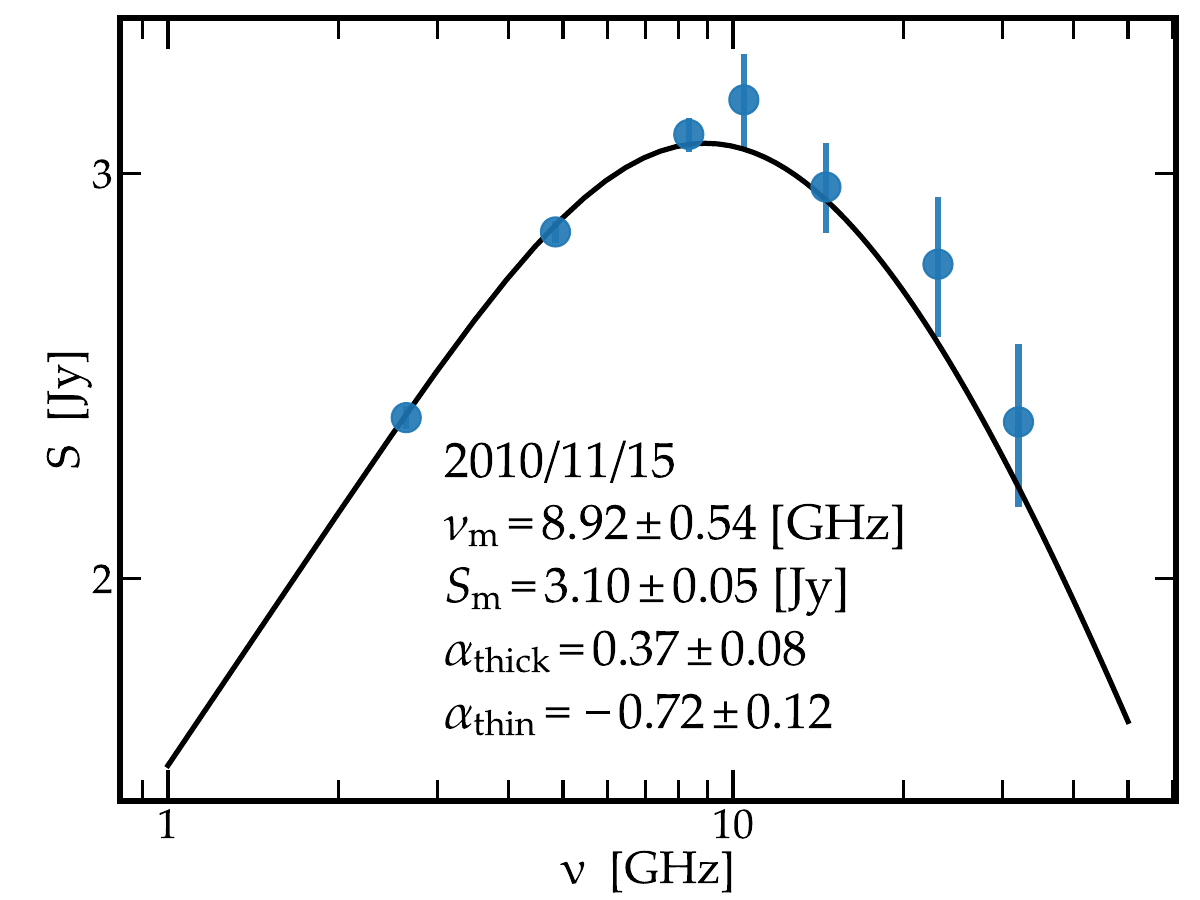}
\includegraphics[width=0.32\textwidth]{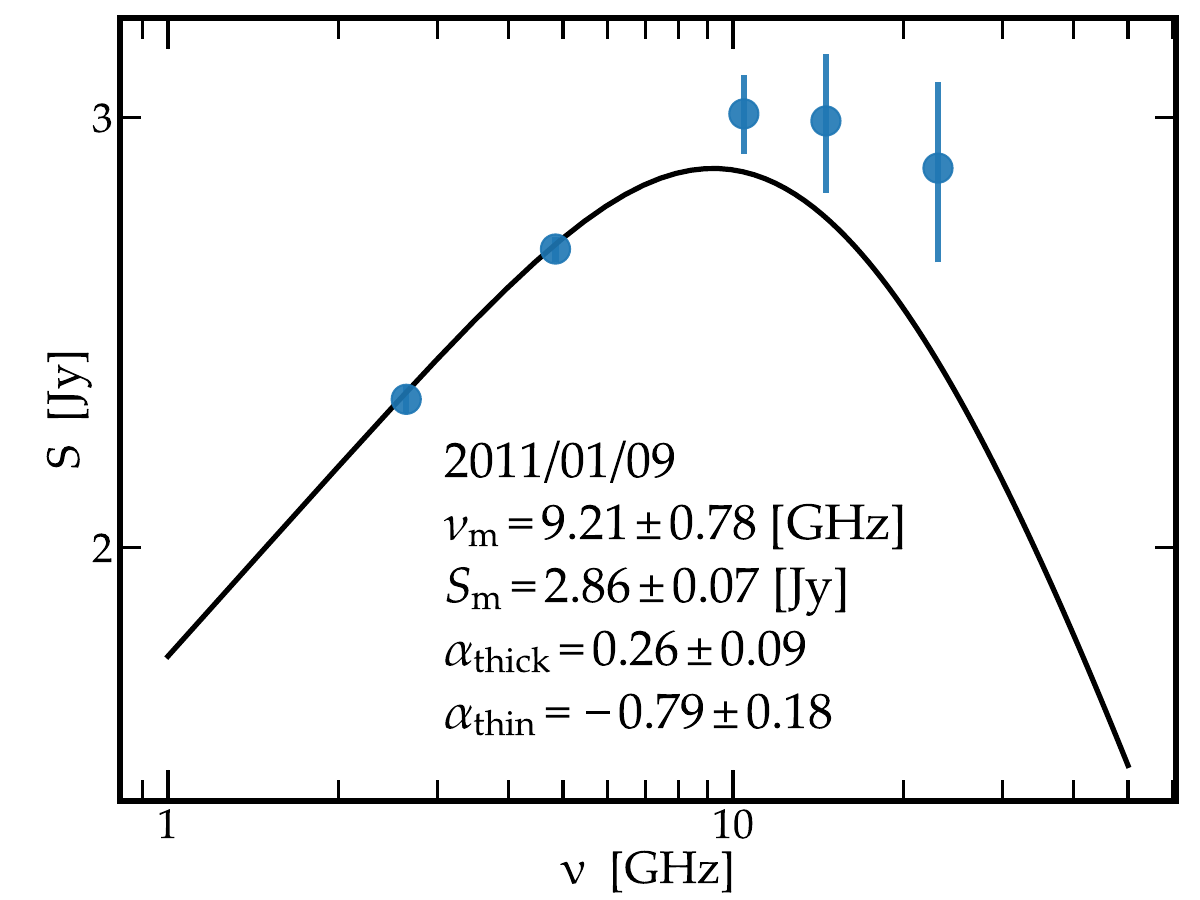}
\end{figure*}
\begin{figure*}[htbp]
\raggedright
\includegraphics[width=0.32\textwidth]{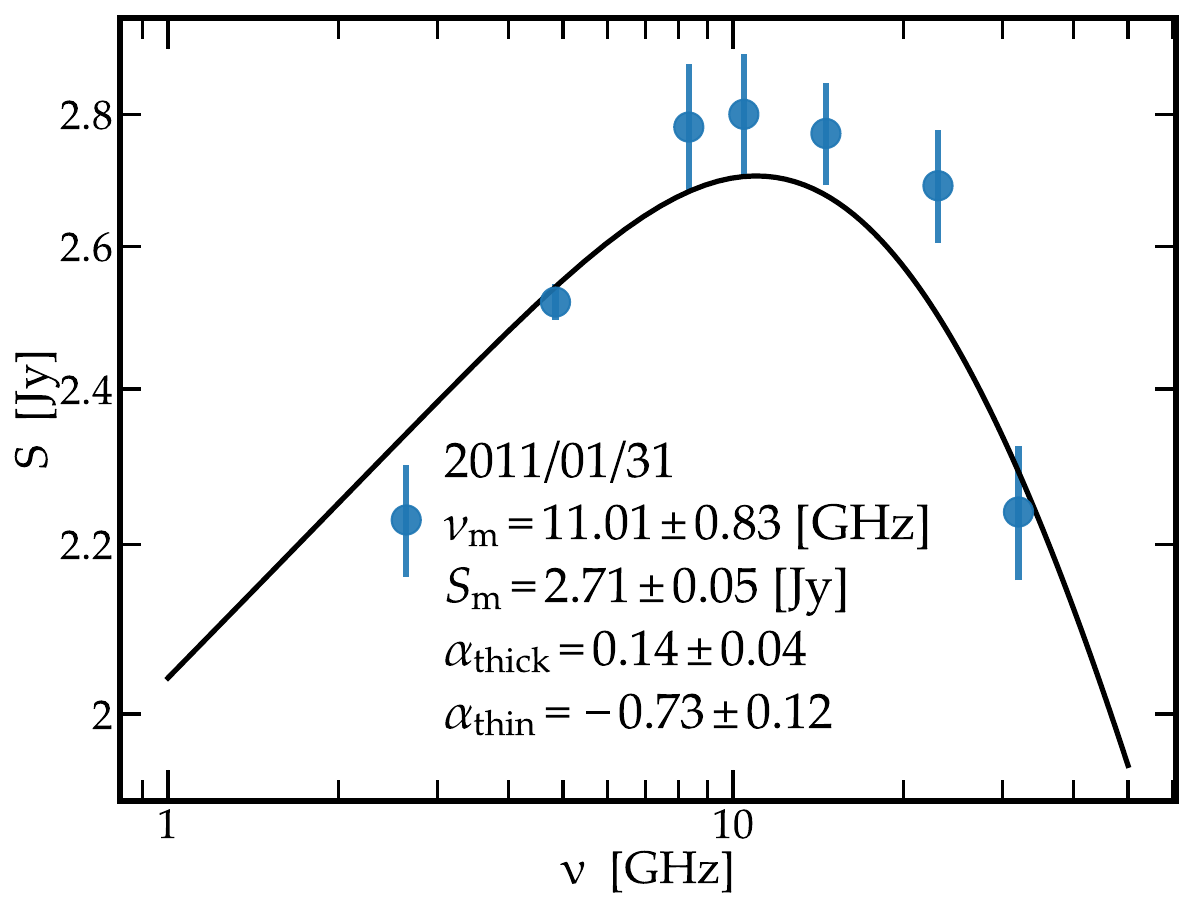}
\includegraphics[width=0.32\textwidth]{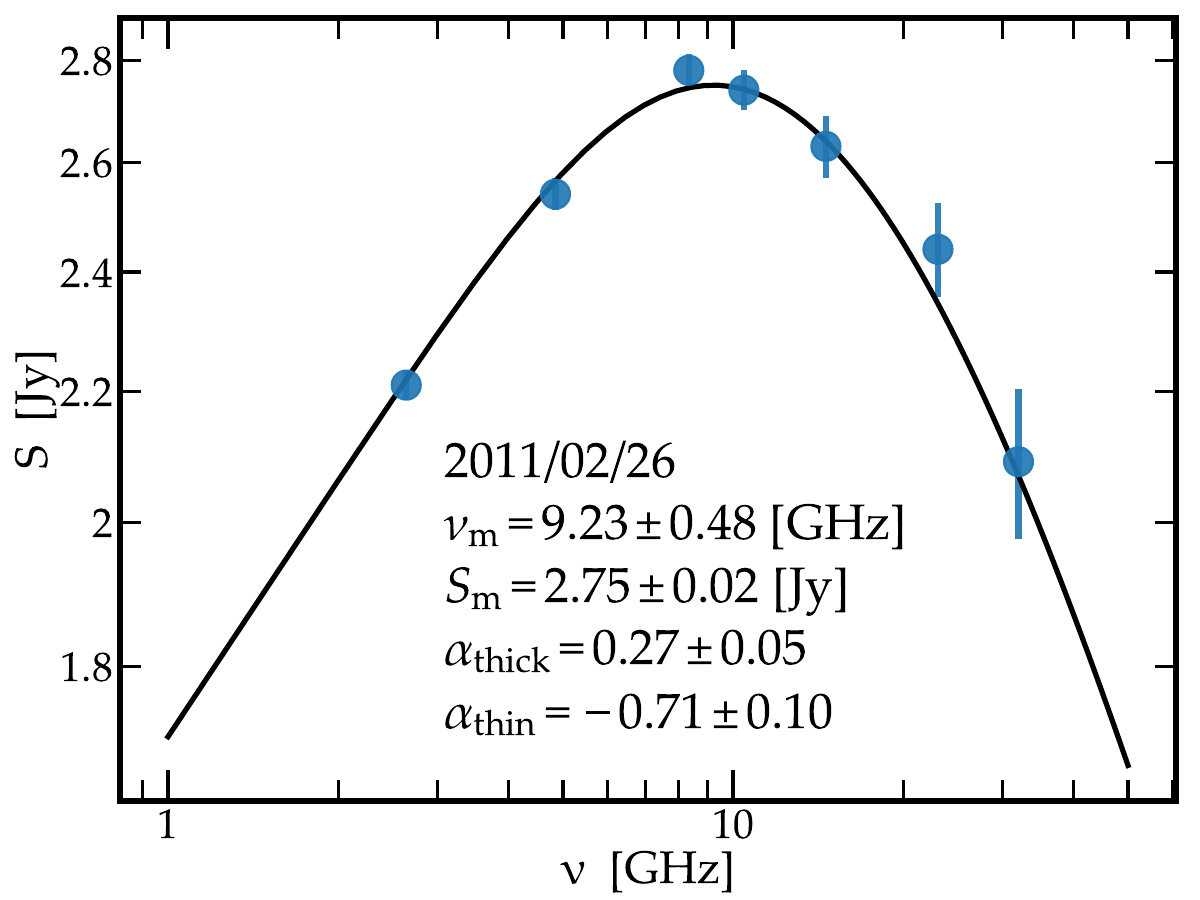}
\includegraphics[width=0.32\textwidth]{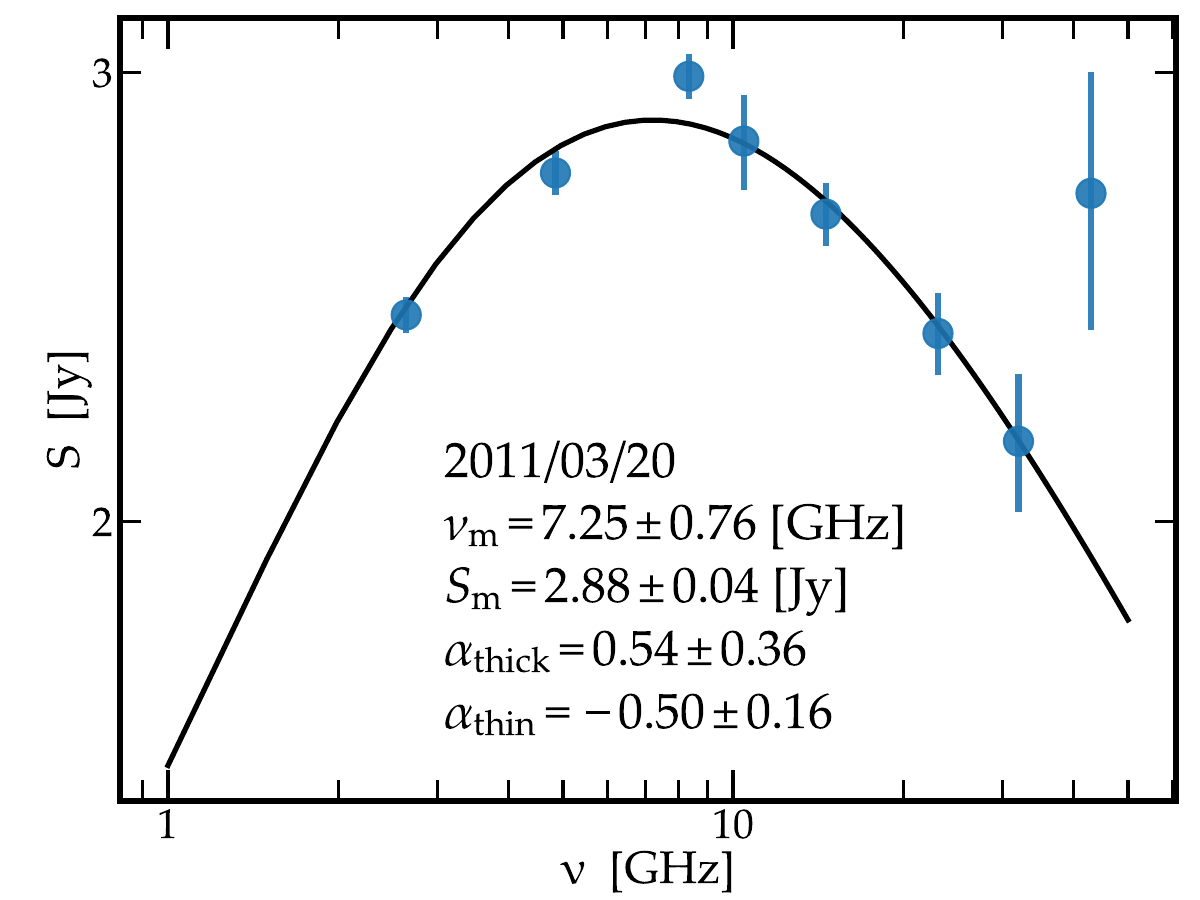}
\includegraphics[width=0.32\textwidth]{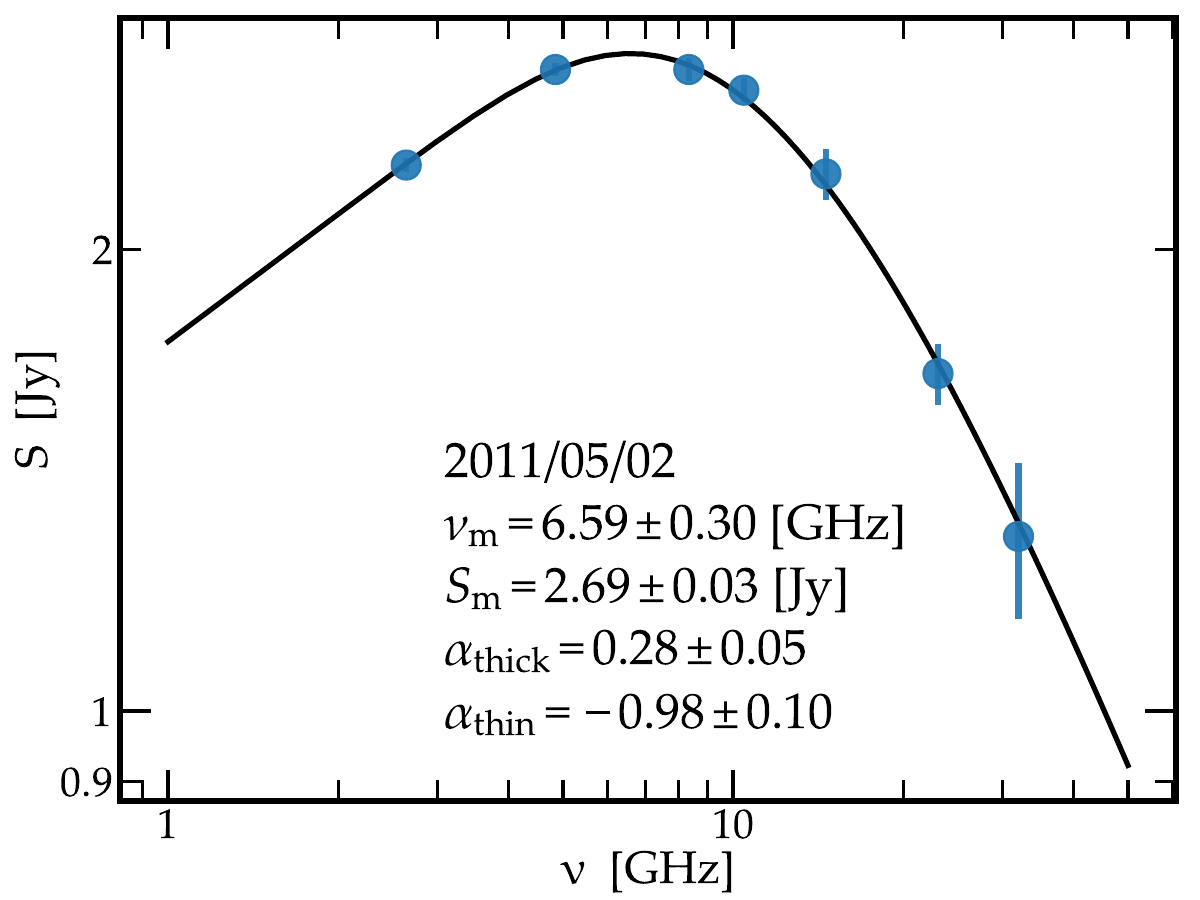}
\includegraphics[width=0.32\textwidth]{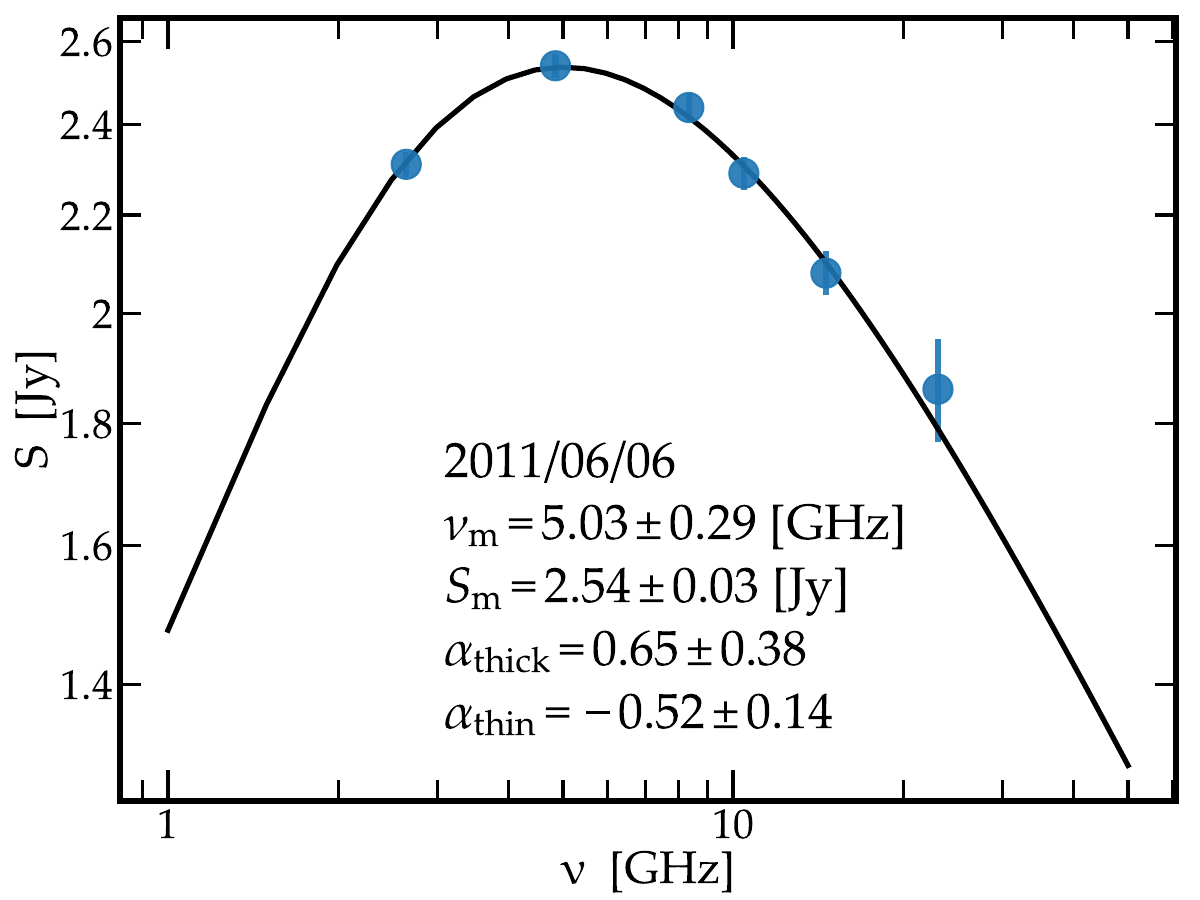}
\caption{SSA spectral fitting results used to estimate the magnetic field strength for blazar 1156+295 from Effelsberg single-dish observations between 2007 and 2012. Only the spectra exhibit peaked shape are presented. Epochs which do not conform to the SSA spectral shape (such as monotonically decreasing, monotonically increasing or concave spectra) are excluded.}
\label{figureD1}
\end{figure*}

%\section{Using Chinese, Japanese, and Korean characters}

%Authors have the option to include names in Chinese, Japanese, or Korean (CJK) characters in addition to the English name. The names will be displayed in parentheses after the English name. The way to do this in AASTeX is to use the CJK package available at \url{https://ctan.org/pkg/cjk?lang=en}. Further details on how to implement this and solutions for common problems, please go to \url{https://journals.aas.org/nonroman/}.

%% For this sample we use BibTeX plus aasjournalv7.bst to generate the
%% the bibliography. The sample7.bib file was populated from ADS. To
%% get the citations to show in the compiled file do the following:
%%
%% pdflatex sample7.tex
%% bibtext sample7
%% pdflatex sample7.tex
%% pdflatex sample7.tex

\bibliography{sample701}{}
\bibliographystyle{aasjournalv7}

%% This command is needed to show the entire author+affiliation list when
%% the collaboration and author truncation commands are used.  It has to
%% go at the end of the manuscript.
%\allauthors

%% Include this line if you are using the \added, \replaced, \deleted
%% commands to see a summary list of all changes at the end of the article.
%\listofchanges

\end{document}